\documentclass[aps,superscriptaddress,double-spaced,floatfix,showkeys,showpacs,longbibliography]{revtex4}
\usepackage{amsmath,amssymb,amsfonts}
\usepackage{float}
\usepackage{hyperref}
\usepackage{graphicx}
\usepackage{xcolor}
\usepackage{tensor}
\usepackage{hyperref} 
\usepackage[normalem]{ulem}
\usepackage{amsmath,amssymb,amsfonts}
\usepackage{float}
\usepackage{hyperref}
\usepackage{graphicx}
\usepackage{xcolor}
\usepackage{tensor}
\usepackage{hyperref}
\usepackage[normalem]{ulem}
\usepackage{color,soul}
\begin{document}
\title{The effects of the QCD critical point on the spectra and flow coefficients of hadrons}
\author{Sushant K. Singh}
\email[Correspondence email address: ]{sushant7557@gmail.com}
\affiliation{Variable Energy Cyclotron Centre, 1/AF, Bidhan Nagar , Kolkata, India}
\affiliation{HBNI, Training School Complex, Anushakti Nagar, Mumbai 400085, India}
\author{Jan-e Alam}
\affiliation{Variable Energy Cyclotron Centre, 1/AF, Bidhan Nagar , Kolkata, India}
\affiliation{HBNI, Training School Complex, Anushakti Nagar, Mumbai 400085, India}
\date{\today}
\begin{abstract}
The space-time evolution of the hot and dense fireball of quarks and gluons produced 
in ultra-relativistic heavy-ion collisions at non-zero baryonic chemical potential and temperature
has been studied by using relativistic viscous causal hydrodynamics.
For this purpose a numerical code has been developed to solve the relativistic viscous causal 
hydrodynamics in (3+1)-dimensions with the inclusion of QCD critical point (CP) through the equation 
of state and scaling behaviour of the transport coefficients.  
We have evaluated the transverse momentum spectra, directed and elliptic flow coefficients of 
pions and protons to comprehend the effect of CP on these observables by using this code. It is found that 
the integration over the entire space-time history of the fireball largely obliterates the effects of 
CP on the spectra and flow coefficients. 
\end{abstract}
\maketitle 
\section{\label{sec:intro} Introduction}
Calculations based on lattice QCD (Quantum Chromodynamics) and effective 
field theoretical models at non-zero temperature ($T$)  
and baryonic chemical potential ($\mu_B$) affirm a complex phase 
diagram~\cite{bazavov,qcd1,qcd2,qcd3,qcd4,qcd5}. 
It is  established by lattice QCD calculations that at high $T$ and low $\mu_B$ $(\rightarrow 0)$  
the quark-hadron transition is a crossover. However,   
the transition may be first order in nature~\cite{qcd4,qcd6} at low $T$ and high $\mu_B$. 
Therefore, it is expected that between the 
crossover and the first order transition there may exist a point 
in the $\mu_B-T$ plane,  called the Critical End Point or simply Critical Point (CP) 
where the first order transition ends and  the cross over begins~\cite{Fodor:2004nz}.  
The location of the CP is not yet precisely known from the lattice QCD 
based calculations~\cite{lqcd} due to
well-known sign problem for spin 1/2 particle (quark) ~\cite{Gavai}
at non-zero $\mu_B$.

However, there are several calculations based on effective field theoretic models 
at non-zero baryon density ~\cite{AB,BJS,TKH,JW} predicting diverse 
locations for the CP in the $\mu_B-T$ plane~\cite{Holography}. The 
coordinates of the CP, ($\mu_{Bc},T_c)$, depend on the values of the parameters of the model.
In the present work we take, $(\mu_{Bc},T_c)=(350 \text{ MeV},143.2 \text{ MeV})$
as a representing point for CP in the QCD phase diagram. 
It is expected that the physical behaviour of
the system reflected through various observables will not depend
on a particular choice of $(\mu_{Bc},T_c)$

It is generally accepted that the quark gluon plasma (QGP) system formed with
small  $\mu_B$ and high $T$ in nuclear collisions
at top Relativistic Heavy Ion Collider (RHIC) and Large Hadron Collider (LHC) energies revert to hadronic phase via cross over transition. The ongoing Beam Energy Scan - II program at RHIC, the upcoming Compressed Baryonic Matter (CBM) and the Nuclotron based Ion Collider fAcility (NICA) experiments 
are projected to produce QGP at high $\mu_B$ and lower $T$ which revert to hadronic phase through a first order phase transition. The QCD matter at different values $T$ and $\mu_B$ can be produced in nuclear collisions by regulating the center of mass energy per nucleon ($\sqrt{s_{NN}}$) of the colliding
nuclei and scanning  through different rapidities at a given colliding energy. Therefore, the colliding energy and the rapidity ($y$) bin should judiciously be chosen to approach the critical point at $(\mu_{Bc},T_c)$. 

Several signatures of the CP has been proposed in the literature. One of the early work~\cite{stephanov} predicts that the existence of the CP will be associated with large event-by-event fluctuation 
of low momentum pions and suppressed fluctuation of $T$ and $\mu_B$. The non-monotonic  dependence of multiplicity fluctuations on $\sqrt{s_{NN}}$~\cite{nxu}, the $y$ dependence of cumulants of the event-by-event proton distributions~\cite{yyin1} 
and the multiplicity fluctuations of pions and protons~\cite{stephanovprl1}, appearance of negative
kurtosis of the order parameter fluctuation~\cite{stephanovprl2} are some of the proposed signals of the CP (for a review see ~\cite{yyin} and references therein). 

Relativistic hydrodynamics (see~\cite{Romatschke:2017ejr,Shen:2020mgh} for review and references therein)
has been used extensively to analyze 
various experimental data originating from heavy ion collisions over a wide range of
$\sqrt{s_{NN}}$ to extract properties of hot and dense QCD matter. It will be interesting
to study how the hydrodynamic evolution of the QCD matter will be affected by
the presence of CP. 
The effects of CP enter into the relativistic viscous hydrodynamics through the EoS and transport 
coefficients.  
The CP changes the EoS and various transport coefficients drastically.
Therefore, it will be useful to examine the effects of CP 
on some of the observables 
{\it i.e.} the transverse momentum ($p_T$) distribution of the 
hadrons, the $p_T$ and $y$ dependence of flow coefficients.

We will use relativistic viscous causal hydrodynamics of
Israel and Stewart with EoS containing the effect of CP and the 
scaling behaviour of shear and bulk viscosities near the CP. 
In a recent study we have found that
the CP has the potential to substantially alter the spin polarization of hadrons~\cite{Singh:2021yba}.
In the present work we will investigate the response of the $p_T$ and $y$
distributions, directed and elliptic flow coefficients  of hadrons (proton and pion) to the QCD
critical point. 

For this exercise, a numerical code in FORTRAN has been 
developed \textcolor{blue}{from the scratch} to solve (3+1)-dimensional viscous relativistic 
causal hydrodynamics using the algorithm detailed in Ref.~\cite{karpenko2014}. 
The code includes the effect of CP through the EoS and the scaling behaviour of the transport
coefficients. The subroutines used for the initial condition and the EoS to solve 
hydrodynamic equations have been extensively tested by reproducing the results 
available in Refs.~\cite{chunshen2020} and ~\cite{parotto2020} respectively. 
The CORNELIUS code~\cite{Cornelius} has been used to find the freeze-out
hypersurface characterized by constant energy density. 
The results from the code have been contrasted with the known analytical results  
of Ref.~\cite{gubser2010}  
and numerical results from the codes  AZHYDRO~\cite{azhydro}, MUSIC~\cite{music} and 
vHLLE~\cite{karpenko2014} by excluding the CP. 

The paper is organized as follows. In section II the method for the numerical solution of hydrodynamic equations has been discussed. Relevant inputs {\it e.g.} the initial condition, equation of state (EoS), transport coefficients are presented through different subsections of this section. We present the results in 
section III and section IV is devoted to summary and discussions.

\section{Numerical Solution of  Relativistic Hydrodynamics} 
Throughout the paper we use natural units 
with $c=\hbar=k_B=1$ where $c$ is the speed of light in vacuum, 
$h$ ($=2\pi\hbar$) is the 
Planck's constant and $k_B$ is the Boltzmann's constant. The flat space time metric is taken as $g_{\mu\nu}=\text{diag}(1,-1,-1,-1)$.
\subsection{\label{sec:hydro} Hydrodynamic Equations}
The relativistic hydrodynamic equations governing the evolution of the system are: 
\begin{align}
\partial _{\mu}T^{\mu \nu}=0 \nonumber \\
\label{charge_conservation}  \partial _{\mu}J_B^{\mu}=0
\end{align}
where $T^{\mu \nu}$ is the energy-momentum tensor and $J_B^{\mu}$ is the net-baryon number current. In the Landau frame, $T^{\mu \nu}$ and $J_B^\mu$ are given by
\begin{align}
T^{\mu \nu} &= \varepsilon \,u^\mu\,u^\nu - (P+\Pi)\Delta^{\mu \nu} +\pi^{\mu \nu}\\
J_B^\mu &= n_B\, u^\mu +V^\mu
\end{align}
where $\varepsilon,\ n_B,\ P$, $u^\mu$, $\Pi$, $\pi^{\mu\nu}$,
$V^\mu$ and  
$\Delta^{\mu \nu} (=g^{\mu \nu}-u^\mu\, u^\nu)$ denote respectively
the energy density, net-baryon number
density, thermodynamic pressure, four-velocity,
bulk pressure, shear-stress tensor,  baryon diffusion four-current and 
projector tensor onto the space orthogonal to $u^\mu$. 
In the Israel-Stewart framework the viscous terms obey relaxation 
type equations which are taken as~\cite{karpenko2014,denicol2018_transcoeff}:
\begin{align}
u^{\gamma}\partial _{\gamma}\Pi &=-\frac{\Pi -\Pi _{NS}}{\tau _{\Pi}}-\frac{4}{3}\Pi\partial _{\gamma}u^{\gamma} \label{eqn:IS_bulk_relax}\\
\langle u^{\gamma}\partial _{\gamma}\pi ^{\mu \nu}\rangle &=-\frac{\pi ^{\mu \nu}-\pi ^{\mu \nu}_{NS}}{\tau _{\pi}}-\frac{4}{3}\pi ^{\mu \nu}\partial _{\gamma}u^{\gamma}\label{eqn:IS_shear_relax}
\\
u^{\gamma}\partial _{\gamma}V^\mu &=-\frac{V^\mu -V^\mu _{NS}}{\tau _{V}}-V^\mu\partial _{\gamma}u^{\gamma}\label{eqn:IS_bardiff_relax}
\end{align}
where $\langle \cdot \rangle$ is defined as
$$\langle A^{\mu \nu} \rangle=\left( \frac{1}{2}\Delta ^{\mu}_{\alpha}\Delta ^{\nu}_{\beta}+\frac{1}{2}\Delta ^{\nu}_{\alpha}\Delta ^{\mu}_{\beta}-\frac{1}{3}\Delta ^{\mu \nu}\Delta _{\alpha \beta}\right)A^{\alpha \beta}$$
and $\Pi_{NS}$, $ \pi_{NS} ^{\mu \nu}$, $V_{NS}^\mu$ are the Navier-Stokes limit of $\Pi$, $ \pi^{\mu \nu}$ and $V^\mu$ respectively and, are given by
\begin{align}
\Pi_{NS} &= -\zeta \theta \\
 \pi_{NS} ^{\mu \nu}&= 2\eta \left\langle\partial ^\alpha u^\beta \right\rangle
\\
V_{NS}^{\mu} &= \kappa_B \Delta^{\mu \nu}\partial_\nu \left( \frac{\mu_B}{T}\right) 
\end{align}
The transport coefficients are positive \emph{i.e.} $\eta ,\zeta, \kappa_B > 0$
where $\eta$, $\zeta$ and $\kappa_B$ are the shear viscosity, bulk viscosity and 
thermal conductivity respectively.
In the present study we take $V^\mu = 0$. As mentioned above
a numerical code has been developed  
in FORTRAN programming language to solve the hydrodynamic equations in Milne 
coordinates $(\tau,x, y, \eta_s)$, using the relativistic HLLE algorithm as 
in~\cite{karpenko2014}, where $\tau=\sqrt{t^2-z^2}$ and 
$\eta_s=\tanh^{-1}(z/t)$. 
The various inputs to the numerical program needed for this study are 
detailed below.

\subsection{\label{sec:ic}Initial Condition}
The Glauber model has been used to estimate the energy 
density profile at the initial time, $\tau_0$, required 
to solve the hydrodynamical equations.
The value of $\tau_0$, for the hydrodynamic simulation is assumed as 
the time taken by the two colliding nuclei to pass through one another 
for $\sqrt{s_{NN}}\,\leq\, 24$ GeV which is determined by the following expression
$$\tau_0 \approx \frac{2R}{\gamma_z v_z},$$
where $\gamma_z=\frac{1}{\sqrt{1-v_z^2}}$ and $v_z=\tanh (y_b)$, 
with $y_b=\cosh^{-1} (\sqrt{s}/2m_N)$ as the beam rapidity, $m_N$ 
is the mass of a nucleon and $R$ is the radius of nucleus. 
The value of $\tau_0$ is taken as 1 fm/c for $\sqrt{s_{NN}} > 24$ GeV. 

The inelastic nucleon-nucleon cross-section ($\sigma^{\texttt{in}}_{NN}$) 
required as an input to the Glauber model as a function of the colliding energy $\sqrt{s}$ (in GeV)
is taken from the following parametrization
~\cite{sigmaNN_parametrization1,sigmaNN_parametrization2}:
\begin{align*}
 \sigma^{\texttt{tot}}_{NN}(\sqrt{s}) &= 42.6s^{-0.46}-33.4s^{-0.545}+35.5\\
 & \qquad +0.307 \ln^2(s/29.1)\\
 \sigma^{\texttt{el}}_{NN}(\sqrt{s}) &= 5.17+12.99s^{-0.41}+0.09\ln^2(s/29.2)\\
 \sigma^{\texttt{in}}_{NN}(\sqrt{s}) &= \sigma^{\texttt{tot}}_{NN} - \sigma^{\texttt{el}}_{NN}
\end{align*}

Our model for the initial condition is based on the inputs taken from~\cite{denicol2018_transcoeff,chunshen2020}. The collision axis is assumed to be along the $z$-axis. 
Let $n_A$ and $n_B$ denote the number of 
wounded nucleons per unit area in the transverse 
plane of the two colliding nuclei $A$ and $B$,  respectively moving along the positive and negative $z$-axis. The  $n_A$ ($n_B$) is given by 
\begin{equation}
n_{A,B}=T_{A,B}\left[ 1-\left( 1-\frac{\sigma^{\texttt{in}}_{NN}T_{B,A}}{N_{B,A}} \right)^{N_{B,A}}\right].
\end{equation}
where $N_A$ and $N_B$ denote the total number of nucleons in $A$ and $B$, respectively, and the thickness functions $T_A(x,y)$ and $T_B(x,y)$ are calculated as follows:
\begin{equation}
T_{A,B}(x,y) = \int_{-\infty}^{\infty}\varrho_{A,B}(x,y,z')\, dz',
\end{equation}
where $\varrho_{A,B}(x,y,z)$ is the nuclear density profile assumed to have the Woods-Saxon shape,
\begin{equation}
\varrho_{A,B}(x,y,z) = \frac{\varrho_0}{1+e^{\frac{r-R(\theta)}{\delta}}}.
\end{equation}
The constant $\varrho_0$ is chosen to satisfy the relation:
\begin{equation}
\int \varrho_i(\vec{r})\, d^3\vec{r} = N_i, \qquad i=A,B
\end{equation}
and $R(\theta)$ has been taken as a function of the polar angle, $\theta$, to account for any deformation of the nucleus, and is given by
\begin{equation}
R(\theta) = R_0\left[ 1+ \beta_2 Y_{2,0}(\theta)+\beta_4 Y_{4,0}(\theta)\right]
\end{equation}
where $Y_{l,m}(\theta,\phi)$ denotes the spherical harmonics. In this study, we consider nuclei to be spherical and take $\beta_2,\ \beta_4=0$. For gold (Au) nucleus, $R_0$ is taken as 6.37 fm. The initial energy density is assumed to have the following form
\begin{equation}
\varepsilon (x,y,\eta_s;\tau_0) = e (x,y) \, f(\eta_s)
\end{equation}
where $f(\eta_s)$ is given by~\cite{chunshen2020}
\begin{align}
f(\eta_s) &= \exp\left[ -\frac{(|\eta_s -y_{\texttt{CM}}|-\eta_0)^2}{2\sigma_\eta^2}\theta(|\eta_s -y_{\texttt{CM}}|-\eta_0)\right].
\end{align}
where
$$y_{\texttt{CM}} = \text{arctanh}\left[ \frac{n_A-n_B}{n_A+n_B}\tanh (y_{b})\right]$$
and $e(x,y) =\mathcal{N}_e M(x,y)$ with  
$$M(x,y)= m_N\sqrt{n_A^2+n_B^2+2n_An_B\cosh (2y_{b})}.$$
The normalization constant $\mathcal{N}_e$ is determined by demanding local 
energy-momentum conservation~\cite{chunshen2020}. The following profile 
for the initial velocity distribution is assumed here,
\begin{equation}
u^\mu (x,y,\eta_s)= \left( \cosh (\eta_s),0,0,\sinh (\eta_s)\right).
\end{equation}
We take the following profile for the initial baryon density,
\begin{equation}
n_B (x,y,\eta_s;\tau_0) = \mathcal{N}_B\left[ g_A(\eta_s)n_A(x,y) + g_B(\eta_s)n_B(x,y)\right]
\end{equation}
where $g_A(\eta_s)$ and $g_B(\eta_s)$ are given by~\cite{denicol2018_transcoeff} 
\begin{align*}
g_A(\eta_s) &= \theta (\eta_s -\eta_{B,0})\exp \left[  -\frac{(\eta_s-\eta_{B,0})^2}{2\sigma _{B,{\texttt{out}}}^2}\right] \\
& \qquad + \theta (\eta_{B,0}-\eta_s)\exp \left[  -\frac{(\eta_s-\eta_{B,0})^2}{2\sigma _{B,{\texttt{in}}}^2}\right]\\
g_B(\eta_s) &= \theta (\eta_s +\eta_{B,0})\exp \left[  -\frac{(\eta_s+\eta_{B,0})^2}{2\sigma _{B,{\texttt{in}}}^2}\right] \\
& \qquad + \theta (-\eta_{B,0}-\eta_s)\exp \left[  -\frac{(\eta_s+\eta_{B,0})^2}{2\sigma _{B,{\texttt{out}}}^2}\right]
\end{align*}
and $\mathcal{N}_B$ is fixed by the condition
\begin{align*}
&\int \, \tau_0 \, dx\, dy\, d\eta_s \ n_B (x,y,\eta_s;\tau_0) = N_{\texttt{part}}\\
&\Rightarrow \quad \int \,  d\eta_s \ n_B (x,y,\eta_s;\tau_0) = \frac{1}{\tau_0}\left[ n_A(x,y)+n_B(x,y)\right]
\end{align*}
which gives
\begin{equation}
\mathcal{N}_B = \frac{1}{\tau_0}\sqrt{\frac{2}{\pi}}\frac{1}{\sigma _{B,{\texttt{in}}}+\sigma _{B,{\texttt{out}}}}.
\end{equation}
The initial condition, hence, is modeled through 6 parameters, 
$(\tau_0,\eta_0,\sigma _{\eta},\eta_{B,0},\sigma _{B,{\texttt{in}}},\sigma _{B,{\texttt{out}}})$ 
which are chosen from Ref~\cite{chunshen2020} for Au+Au collisions. 
\subsection{\label{sec:eos} Equation of State}
The EoS has been obtained by following the procedure detailed in
Ref~\cite{parotto2020} by assuming that the CP in QCD 
belongs to the same universality class as that of 
3D Ising model. The procedure is briefly reviewed below. 
The pressure at non-zero $T$ and $\mu_B$ can be obtained through a Taylor 
series expansion about $\mu_B = 0$ as follows
\begin{equation}
\label{eqn:tylr_exp}
P_{\text{QCD}} (T,\mu_B) = T^4 \sum_n c_{2n} (T) \left( \frac{\mu_B}{T} \right)^{2n} ,
\end{equation}
where
\begin{equation}
\label{eqn:tylr_coeff}
c_n (T) = \frac{1}{n!} \left. \frac{\partial^n (P/T^4)}{\partial (\mu_B/T)^n} \right|_{\mu_B=0} = \frac{1}{n!} \chi_n (T) \, \, .
\end{equation}
If there were no singularity, then the series expansion  given in 
Eq.(\ref{eqn:tylr_exp}) would have been valid throughout the QCD phase diagram. However, the presence of CP makes some of the coefficients diverge. Hence the pressure can be written as a sum of a regular and a singular part. Equivalently, the expansion coefficients in Eq.(\ref{eqn:tylr_exp}) are replaced by
\begin{equation} 
\label{eqn:tylr_coeff_transf}
T^4 c_n (T) \rightarrow T^4 c_n^{\text{Non-Ising}} (T) + f(T,\mu_B) c_n^{\text{Ising}} (T) .
\end{equation}
where the superscript ``Non-Ising" and "Ising" represent the regular and 
the singular (or critical) contributions respectively. 
$f(T,\mu_B)$ is chosen so that it does not add any other singularity in the problem, and can simply be chosen as
$$f(T,\mu_B)=T_c^4.$$
The critical part is obtained by using the 3D Ising model because the QCD critical point belongs to the same universality class as the 3D Ising model. Hence, the two models must show the same scaling behavior in the critical region. The critical exponents of the 3D Ising model are known through numerical simulations. Hence, by mapping the parameters of the two systems in the critical region, it is possible to extract the critical behavior of QCD near $T_c$. The mapping from the Ising model phase diagram $(r,h)$ to the QCD phase diagram $(\mu_B,T)$ is done 
with the help of the following relations:
\begin{align}
\frac{T - T_C}{T_C} &=  w \left( r \rho \,  \sin \alpha_1  + h \, \sin \alpha_2 \right) \, \, , \nonumber \\ 
\frac{\mu_B - \mu_{BC}}{T_C} &=  w \left( - r \rho \, \cos \alpha_1 - h \, \cos \alpha_2 \right) \label{eqn:map_rh_muT} ,
\end{align} 
The Ising pressure in the critical region is given by
\begin{equation}
P_{\texttt{Ising}}(R,\theta) = h_0M_0 R^{2-\alpha}\left[ \theta \tilde{h}(\theta)-g(\theta)\right]
\end{equation}
where 
$\tilde{h}(\theta) = \theta (1+a\theta ^2+b\theta^4)$, $g(\theta )
=c_0+c_1(1-\theta^2)+c_2(1-\theta^2)^2+c_3(1-\theta^2)^3$;
$h_0,\, M_0,\, a,\, b,\, c_0,\, c_1,\, c_2,\,c_3$ are constants, some given in terms 
of critical exponents. $(R,\theta)$ are related to $(r,h)$ through the following transformations
\begin{align}
h &= h_0 R^{\beta \delta}\tilde{h}(\theta) \label{eqn:hvsRTh}\\
r &= R(1-\theta^2) \label{eqn:rvsRTh}
\end{align} 
where $\beta$ and $\delta$ are critical exponents. Hence the coefficients contributing to the critical part are determined through
\begin{equation}
c_n^{\text{Ising}} (T) = \frac{1}{n!} T^n \left. \frac{\partial^n P^{\text{Ising}}}{\partial \mu_B^n} \right|_{\mu_B=0}  ,
\end{equation} 
The ``Non-Ising'' coefficients are chosen in such a way that at $\mu_B=0$, the expansion coefficients so obtained match the results obtained from LQCD \emph{i.e.}
\begin{equation} 
\label{eqn:NI_coeff}
T^4 c^{\text{LAT}}_n (T) = T^4 c_n^{\text{Non-Ising}} (T) + f(T,\mu_B=0) c_n^{\text{Ising}} (T) .
\end{equation}
Having determined the ``Ising'' and ``Non-Ising'' coefficients, the full pressure is then obtained as
\begin{align} 
\label{eqn:Pfull}
P_{\text{QCD}} (T, \mu_B) &= T^4 \sum_n c_{2n}^{\text{Non-Ising}} (T) \left( \frac{\mu_B}{T} \right)^{2n}\nonumber\\
&+T_c^4\, P^{\text{Ising}} (R(T, \mu_B) ,\theta (T, \mu_B)) . 
\end{align}
The procedure as detailed above gives various thermodynamic observables as a function of $T$ and $\mu_B$. However, for our numerical hydrodynamic code we require pressure $(P)$ as a function of energy density $(\varepsilon)$ and baryon number density $(n_B)$. To construct such a table, we follow a procedure similar to~\cite{rischke_eos}. The $\varepsilon -n_B$ plane is discretized with the following scheme,
$$\Delta \varepsilon \, (\text{GeV/fm}^3) = \left\{ \begin{array}{cc}
                                0.002 & \text{ if } \qquad 0.001 \le \varepsilon < 1.001 \\
                                0.02 & \text{ if }\qquad 1.001 \le \varepsilon < 11.001 \\
                                0.1 & \text{ if } \qquad 11.001 \le \varepsilon < 61.001 \\
                                0.5 & \text{ if } \qquad 61.001 \le \varepsilon < 101.001
                               \end{array}
\right.$$
$$\Delta n_B \, (\text{fm}^{-3})= \left\{ \begin{array}{cc}
                                0.0005 & \text{ if } \qquad 0 \le n_B < 0.15 \\
                                0.001 & \text{ if }\qquad 0.15 \le n_B < 0.3 \\
                                0.01 & \text{ if } \qquad 0.3 \le n_B < 1 \\
                                0.025 & \text{ if } \qquad 1 \le n_B < 5
                               \end{array}
\right.$$
Small value for discretization is chosen to correctly reproduce the critical behavior and discontinuity in thermodynamic quantities due to the first-order transition beyond the critical point. For other values of $(\varepsilon ,n_B)$, the pressure and other thermodynamic variables are obtained through 2D linear interpolation. For all those $(\varepsilon ,n_B)$ where $T$ and $\mu_B$ lies outside the range $(5,450)\text{ MeV}$ and $(0,450)\text{ MeV}$ respectively, the thermodynamic variables are put to zero.
\subsection{\label{sec:transcoeff} Transport Coefficients}
Before providing expressions for the transport coefficients we shall give an account of how the equilibrium 
correlation length can be obtained with an equation of state near the CP. 
The procedure of Ref.\citep{monnaibulk2017} has been followed 
to calculate the equilibrium correlation length $\xi$. 
$\xi$ is computed in the Ising model by taking derivative of equilibrium magnetization, $M(r,h)$, with respect to $h$ at fixed $r$ as follows
$$\xi^2 = \frac{1}{H_0}\left( \frac{\partial M(r,h)}{\partial h}\right)_r$$
where $H_0$ is a dimensionful parameter to get the correct dimensions of $\xi$. We shall take $H_0=1$ 
in our calculations. In fact, the derivative of $M(r,h)$ 
with respect to  $h$ is the 
\begin{figure}[H]
\centering
\includegraphics[height=50mm,width=70mm]{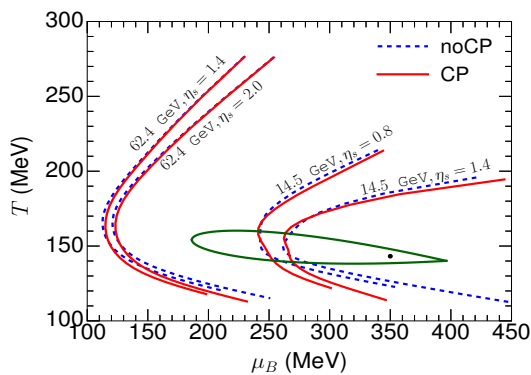}  
\caption{Trajectory traced by the fluid cell at $x=y=0$ in the $T-\mu_B$ plane for different space-time rapidities, two colliding energies (14.5 GeV and 62.4 GeV) and for impact parameter $b=5$ fm. The critical point is indicated in solid black dot at $(T,\mu_B)=(350,143.2)$ MeV and the boundary of the critical region is indicated by the solid green line. }
\label{fig:evln_traj}
\end{figure}
\noindent magnetic susceptibility ($\chi_M$) in the 
Ising model which, near a critical point, goes as $\xi^{2-\eta^\prime}$, where
the value of $\eta^\prime$ is found to be small ($\eta^\prime \approx 0.036$).
In this work we have taken $\eta^\prime=0$. 
The equilibrium magnetization is parametrized in terms of variables $R$ and $\theta$ as
$$M(R,\theta) = M_0R^\beta \theta$$
where $R$ and $\theta$ are related to $r$ and $h$ through Eqns.(\ref{eqn:hvsRTh}) and (\ref{eqn:rvsRTh}). Now using the identity
$$\left( \frac{\partial M}{\partial h}\right)_r = \left( \frac{\partial M}{\partial R}\right)_\theta \left( \frac{\partial R}{\partial h}\right)_r+\left( \frac{\partial M}{\partial \theta}\right)_R\left( \frac{\partial \theta}{\partial h}\right)_r$$
and the expressions for $\left( \frac{\partial R}{\partial h}\right)_r$ and $\left( \frac{\partial \theta}{\partial h}\right)_r$ given in Ref.\cite{parotto2020}, we have
\begin{equation}
\label{eqn:correl_length}
\xi^2 = \frac{M_0}{h_0} \frac{ R^{\beta (1-\delta)}}{2\beta \delta \theta \tilde{h}(\theta)+(1-\theta^2)\tilde{h}'(\theta)} \left[ 1+ (2\beta -1)\theta^2\right]
\end{equation}
Near the critical point, the transport coefficients are expected to vary 
with the correlation length as follows
$$\zeta \sim \xi^3 \quad , \quad \eta \sim \xi^{0.05} \quad , \quad \kappa_T \sim \xi$$ 
We define the region in the $\mu_B-T$ plane bounded by the curve $\xi (\mu_B,T)=\xi_0$ as the critical region \emph{i.e.} for $\xi < \xi_0$, the transport coefficients are regular functions of $T$ and $\mu_B$ but for $\xi > \xi_0$, the transport coefficients must satisfy the scaling laws as defined above. We choose $\xi_0=1.75$ fm. In this work, we only consider 
bulk viscosity ($\zeta$) and shear viscosity ($\eta$). 
The critical behavior of these transport coefficients can then be modeled as
\begin{equation}
\label{eqn:crit_zeta}
\zeta = \zeta_0 \left( \frac{\xi}{\xi_0}\right)^3 \ , \ \eta = \eta_0 \left( \frac{\xi}{\xi_0}\right)^{0.05}
\end{equation}
where $\zeta_0,\ \eta_0$ denote the values outside the critical region which are chosen as\cite{denicol2018_transcoeff,denicol2014_transcoeff}
\begin{align*}
\eta_0 (\mu_B,T) &= 0.08 \left( \frac{\varepsilon + p}{T}\right) \\
\zeta_0 (\mu_B,T) &= 15 \ \eta_0 (\mu_B,T)\left( \frac{1}{3}-c_s^2\right)^2
\end{align*}
The relaxation times in Eqs.(\ref{eqn:IS_bulk_relax}) and (\ref{eqn:IS_shear_relax}) also diverge near the critical point. This is included by using the following expressions for the relaxation times
\begin{equation}
\label{eqn:crit_relax}
\tau_\pi = \tau^0_\pi \left( \frac{\xi}{\xi_0}\right)^{0.05} \quad ,\quad \tau_\Pi = \tau^0_\Pi \left( \frac{\xi}{\xi_0}\right)^3.
\end{equation}
where $\tau^0_\pi$ and $\tau^0_\Pi$ are the relaxation times outside the critical region, which we take as follows (with $C_\eta =0.08$)~\cite{denicol2014},
$$\frac{\tau^0_\pi}{5} =\tau^0_\Pi =\frac{C_\eta}{T}.$$
\begin{figure}[H]
\centering
\includegraphics[scale=0.5]{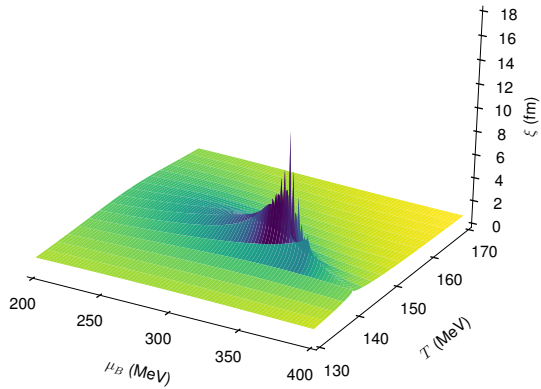}  
\caption{Correlation length, $\xi$, plotted as a function of $\mu_B$ and $T$.}
\label{fig:correl3d}
\end{figure}
\begin{figure*}
\centering
\begin{tabular}{cc}
\includegraphics[height=50mm,width=70mm]{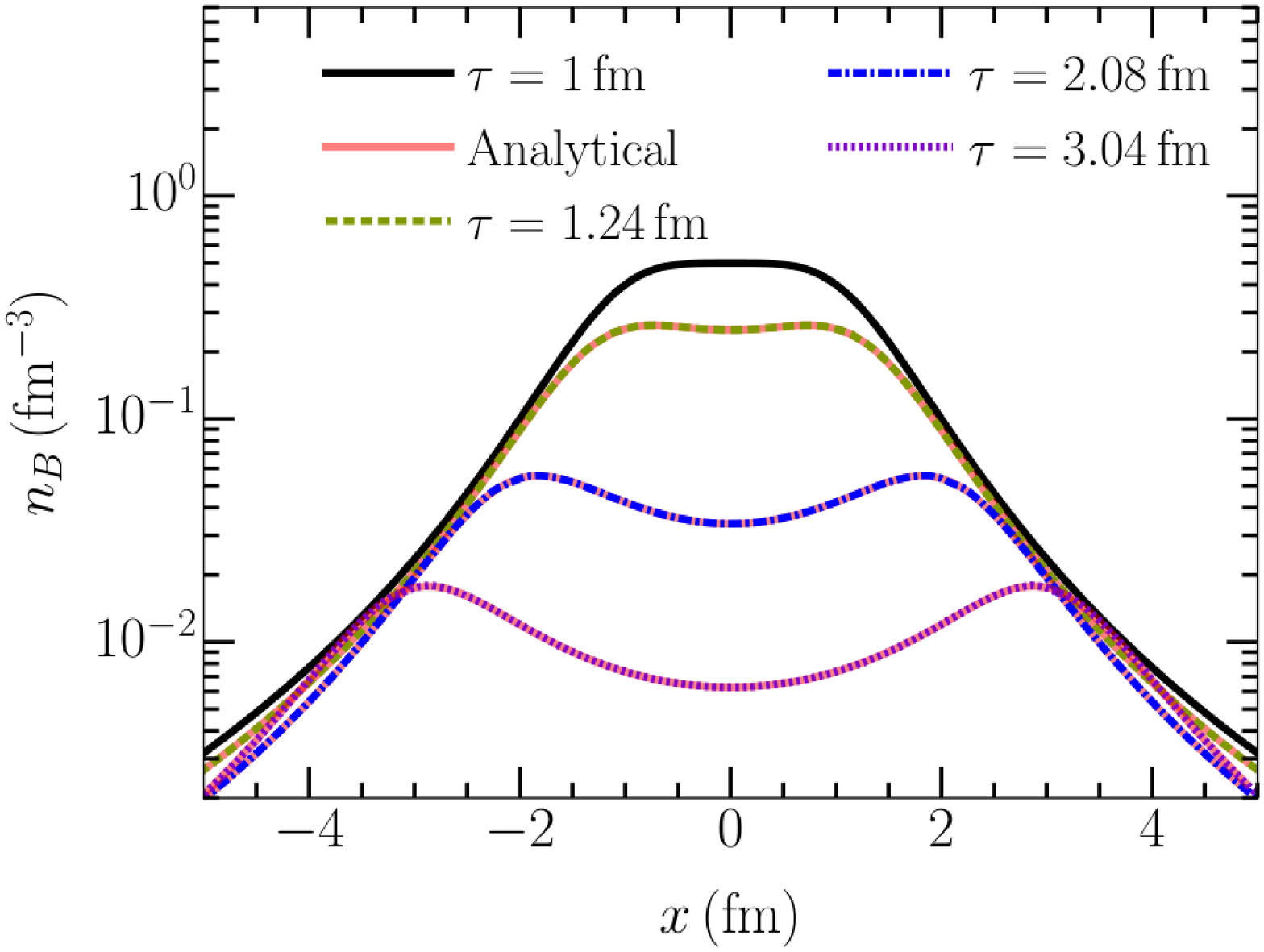} 
\includegraphics[height=50mm,width=70mm]{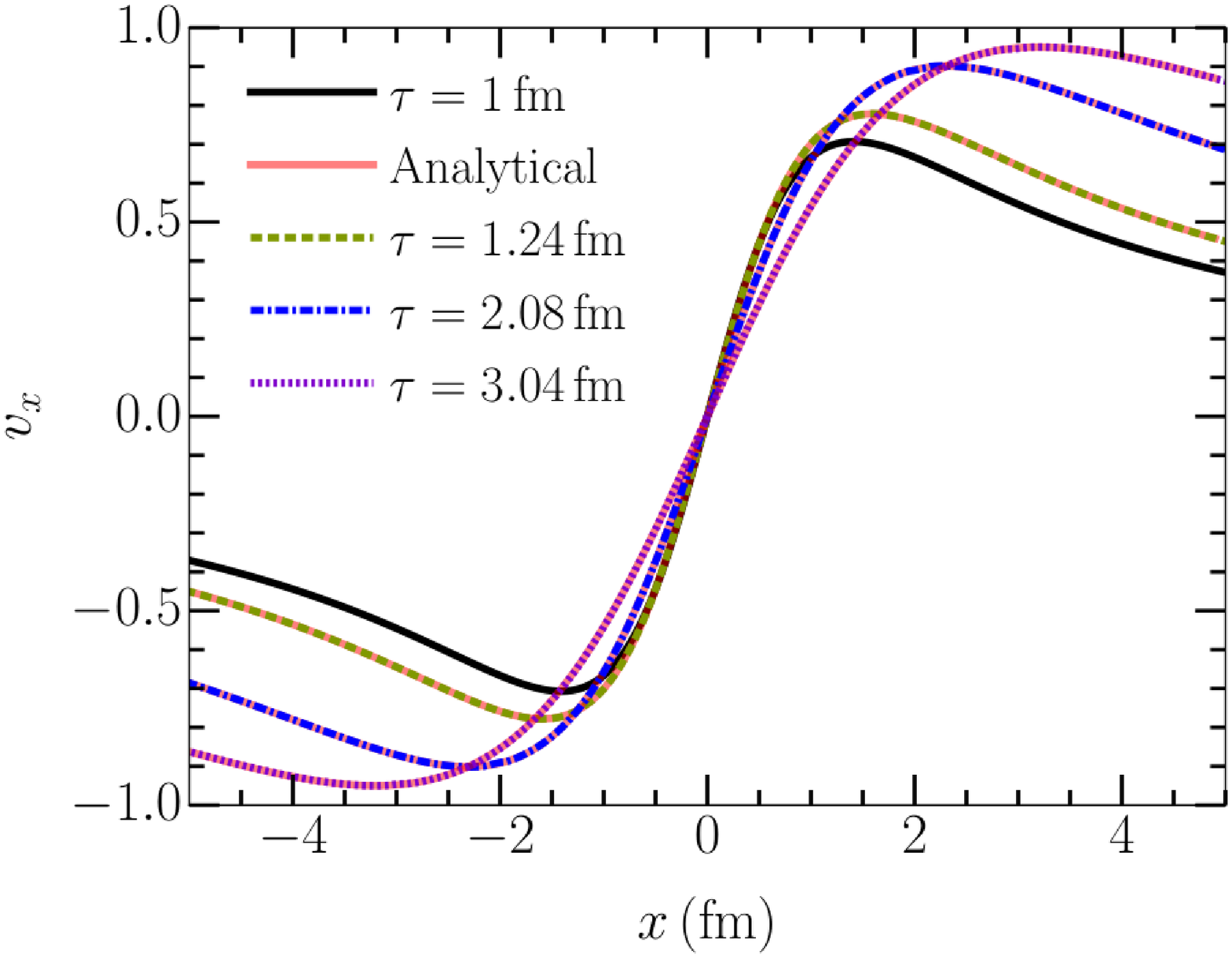} 
\end{tabular}
\caption{Comparison of the code output with analytic Gubser solution. The results are plotted at $y=0$ after setting $q=1$ in Eqs.(\ref{eqn:gub_nB}) and (\ref{eqn:gub_vx}).}
\label{fig:gub}
\end{figure*}

The critical domain is delineated in $\mu_B-T$ plane by 
setting $\xi(\mu_B,T)=\xi_0=1.75$ fm.
The critical point indicated by black dot
and the critical domain enclosed by the green line are shown in
Fig.(\ref{fig:evln_traj}).  The
trajectories of the system in $\mu_B-T$ plane formed at two $\sqrt{s_{NN}}$ 
with and without CP have  been displayed.  
The trajectories of the fluid cells at lower $\sqrt{s_{NN}}$ 
pass through the critical region but for higher $\sqrt{s_{NN}}$ 
 the trajectories remain away from the 
critical region. But for a fixed collision energy, the fluid cell at larger 
space-time rapidity, 
$\eta_s$, lies closer to the critical point and will feel the effect of EoS and enhanced 
values of transport coefficients more.  
The evolution of fluid cell at $(x,y,\eta_s)=(0,0,1.4)$ at $\sqrt{s_{NN}}=14.5$ GeV is displayed in Fig.(\ref{fig:evln_traj}). The results clearly show that the trajectories at higher $\eta_s$  get attracted towards the critical point. 

The correlation length as obtained in Eq.(\ref{eqn:correl_length}) is shown in Fig.(\ref{fig:correl3d}) as function of $\mu_B$ and $T$.  
The drastic increase in $\xi$ in the vicinity of the CP is conspicuous due to the divergent nature of the transport coefficients.
\subsection{Numerical implementation}
We perform the hydrodynamical simulation on a $201\times 201\times 71$ space grid such that $\Delta x=\Delta y=\Delta \eta =0.2 \text{ fm}$. Also time step for the evolution is chosen as $\Delta \tau =0.05 \text{ fm}$. This choice satisfies the CFL (Courant-Friedrichs-Lewy) criterion for the stability of the code. Further the time evolved quantities are written to a data file after evolving for $0.5 \text{ fm}$ time from the previous step. 
We use CORNELIUS~\cite{Cornelius} code to find a constant energy density surface within a computational 
(fluid) cell. The CORNELIUS code provides the coordinates $(\tau_f,x_f,y_f,\eta_{sf})$ and area elements 
$d\Sigma _\nu$ of the freeze-out surface. The quantities like $T$, $\varepsilon$ etc at 
$(\tau_f,x_f,y_f,\eta_{sf})$ are calculated through 2D linear interpolation using the values at the corners of the cell. We shall analyze the effects of the CP on the surface $\varepsilon=0.3 \ \text{GeV/fm}^3$, henceforth denoted as $\Sigma_{CFO}$. 
The aim here is not to compare and reproduce the experimental data but to pinpoint the effects 
that CP will induce on various hydrodynamic quantities and hence on the experimental  observables.

We calculate the space averaged quantity, say transverse velocity,
$v_T=\sqrt{v_x^2+v_y^2}$  by using the following expression:
$$\left\langle v_T \right\rangle (\tau) = \dfrac{\int d^3x \ \varepsilon (\tau,x,y,\eta_s) v_T(\tau,x,y,\eta_s)}{\int d^3x \ \varepsilon (\tau,x,y,\eta_s)}$$. 
where $\epsilon(\tau,x,y,\eta_s)$ is the energy density.
The $p_T$ spectrum of the hadrons,
$\frac{dN}{d^2p_Tdy}$ can be calculated by using the Cooper-Frye formula as:
$$\frac{dN_i}{d^2p_Tdy} = \frac{g_i}{(2\pi)^3}\int d\sigma _{\mu}p^{\mu}f_i(x,p)$$
where $f_i(x,p)$ is given by
$$f_i(x,p)=\frac{1}{e^{(p^{\mu}u_{\mu}-\mu _{B,F})/T_{F}}+a_i}$$
where $T_F$ and $\mu_{BF}$ denote the temperature and chemical potential on the surface $\Sigma_{CFO}$, $a_i=-1$ for bosonic statistics, $a_i=+1$ 
for fermionic statistics and $a_i=0$ for classical (Boltzmann) statistics. 
We do not consider a viscous correction $(\delta f_i)$ to the distribution 
$f_i$ in evaluating $\frac{dN_i}{d^2p_Tdy}$ as the effect 
may not be significant~\cite{monnaibulk2017}. 

The other experimental measurable quantities, like flow coefficients 
are calculated by using the following expression,
\begin{equation}
v_n(p_T,y) = \dfrac{\int d\phi\ \frac{dN}{d^2p_Tdy}\cos (n \phi)}{\int d\phi\ \frac{dN}{d^2p_Tdy}}
\end{equation}
The $p_T$-integrated flow coefficient is obtained as
$$v_n(y) = \int_{p_{T,min}}^{p_{T,max}} dp_T\ v_n(p_T,y)$$
where we take $p_{T,min}=0.2$ GeV and $p_{T,max}=3$ GeV. Similarly, the rapidity integrated flow coefficient is obtained as
$$v_n(p_T) = \int_{y_{min}}^{y_{max}} dy\ v_n(p_T,y)$$
where we have chosen $y_{min}=-1$ and $y_{max}=1$.
\begin{figure*}
\centering
\includegraphics[height=50mm,width=75mm,angle=0]{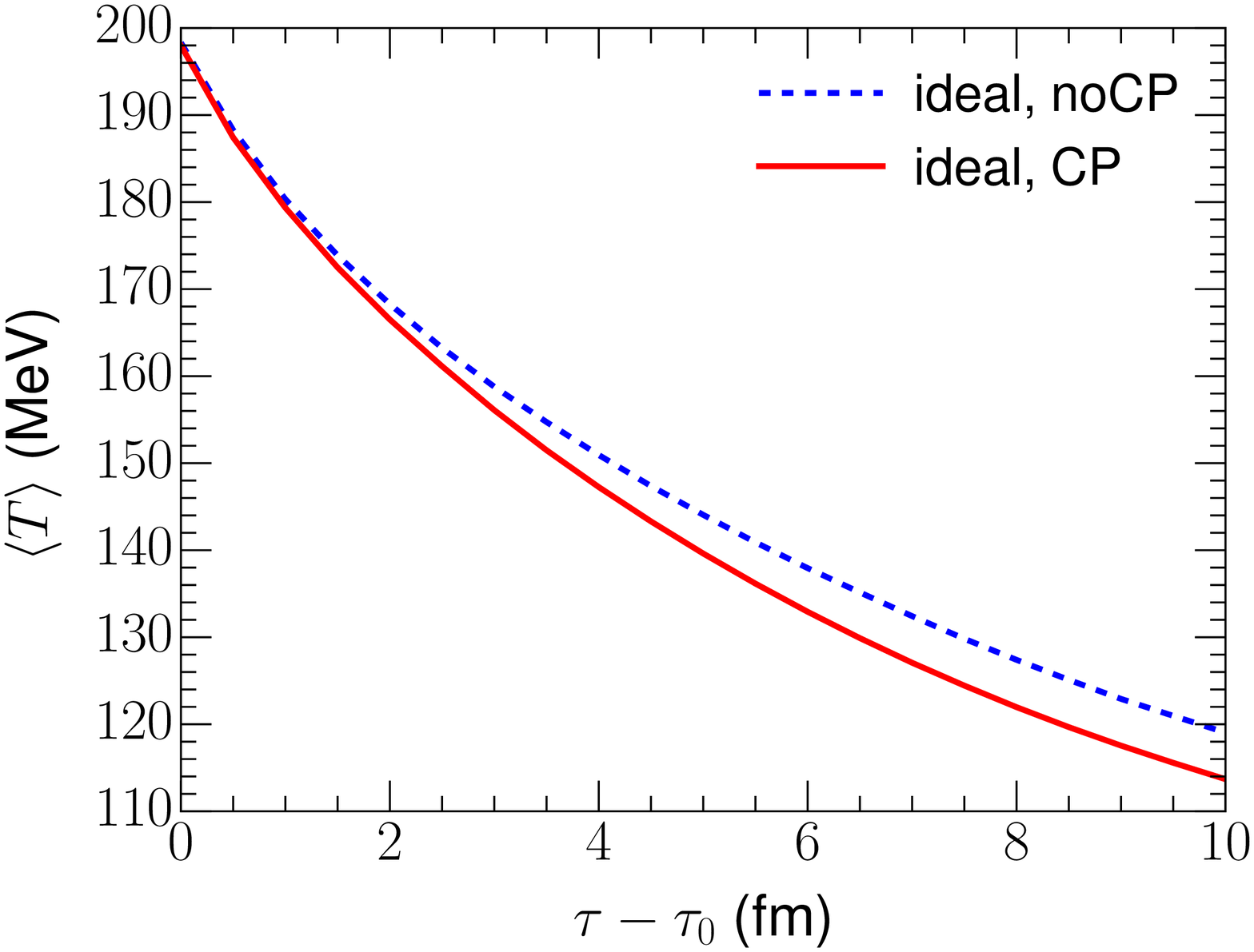}
\includegraphics[height=50mm,width=75mm,angle=0]{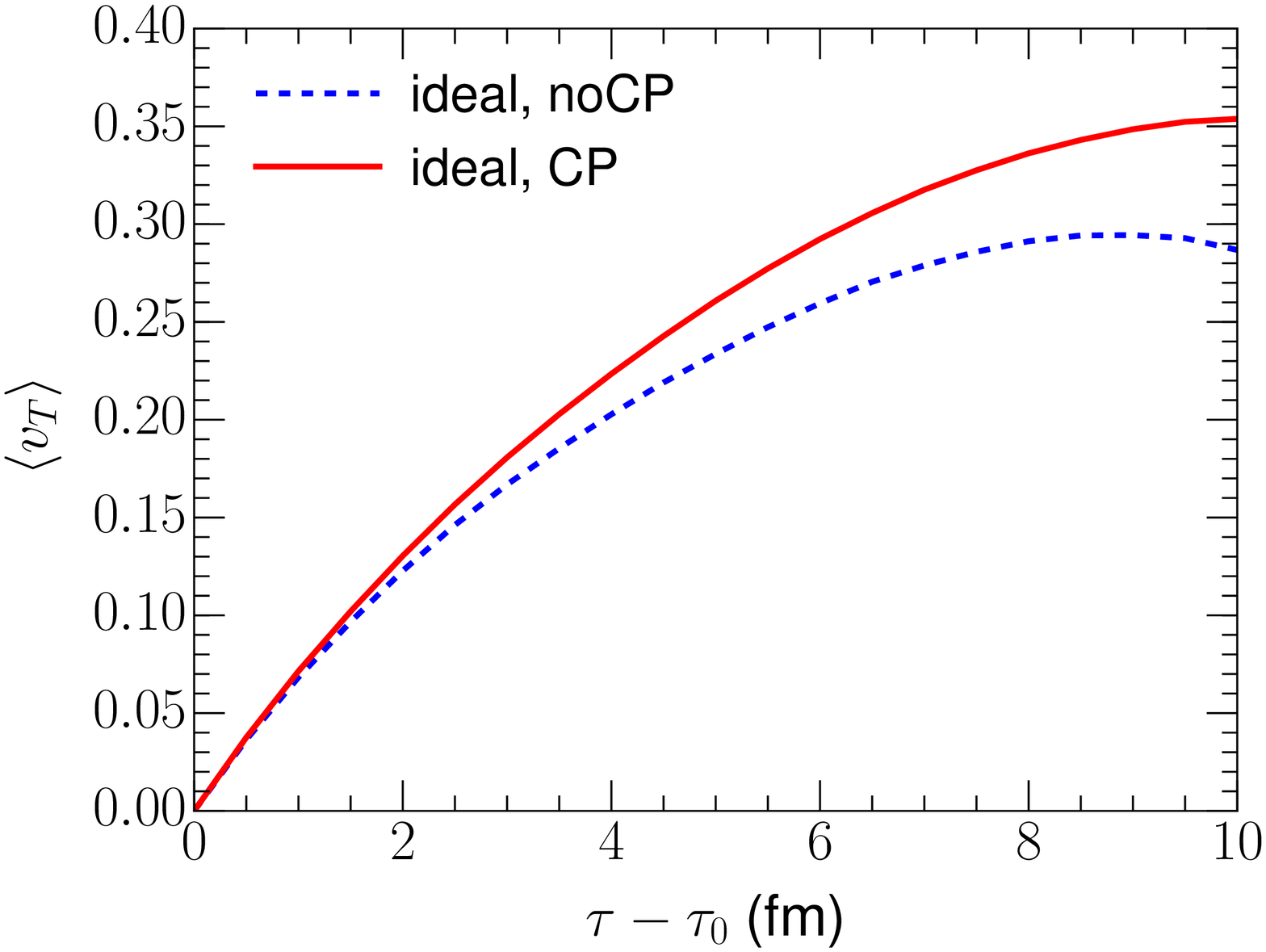} 
\caption{Time evolution of (left) average temperature, and (right) transverse velocity using ideal hydrodynamics and EoS with (CP, red solid line) and without (noCP, blue dashed line) the critical point for Au+Au collision at $\sqrt{s_{NN}}=14.5$ and impact parameter $b=5$ fm.}
\label{fig:avgtemp}
\end{figure*}

\section{Results and Discussions}
We compare numerical results from our code and the analytical Gubser 
solution in Fig.(\ref{fig:gub}). The Gubser solution is given 
by~\cite{gubser2010,denicol2018_transcoeff}
\begin{align}
 \varepsilon (\tau,r) &= \frac{\varepsilon_0}{\tau^4}\frac{(2\, q\, \tau)^{8/3}}{\left[ 1+2 \left(\tau ^2+r^2\right)+\left(\tau ^2-r^2\right)^2 \right]^{4/3}}\label{eqn:gub_ed}\\
 n_B(\tau,r) &= \frac{n_{B0}}{\tau^3}\frac{(2\, q\, \tau)^{2}}{ \left[1+2 \left(\tau ^2+r^2\right)+\left(\tau ^2-r^2\right)^2\right]} \label{eqn:gub_nB}\\
 v_x(\tau,r)&= \frac{2 q^2\tau \, x}{1+ q^2r^2 + q^2\tau ^2}\label{eqn:gub_vx}\\
 v_y(\tau,r)&=\frac{2 q^2\tau \, y}{1+ q^2r^2 + q^2\tau ^2}\label{eqn:gub_vy}
\end{align}
where $r=\sqrt{x^2+y^2}$. The numerical solution is in perfect agreement with the analytical result. 
Next we show the time evolution of the average temperature of the fireball formed in Au+Au collisions 
($\sqrt{s_{NN}}=14.5$ GeV at impact parameter $b=5$ fm) in Fig.(\ref{fig:avgtemp}) 
using the ideal hydrodynamics (all transport coefficients set to zero) and EoS
with (red solid line) and without (blue dashed line) the critical point.  
For a given initial condition, the addition of the critical part to the regular pressure leads to a larger value of the net pressure, and hence a large gradient of the pressure with respect to the vacuum outside. This leads to a faster expansion and a faster rate of cooling as shown in the left panel of Fig.(\ref{fig:avgtemp}). Higher pressure introduced by the CP lead to higher flow too as reflected in $v_T$, as shown in the right panel of Fig.(\ref{fig:avgtemp}). 

\begin{figure}[H]
\centering
\includegraphics[height=48mm,width=75mm,angle=0]{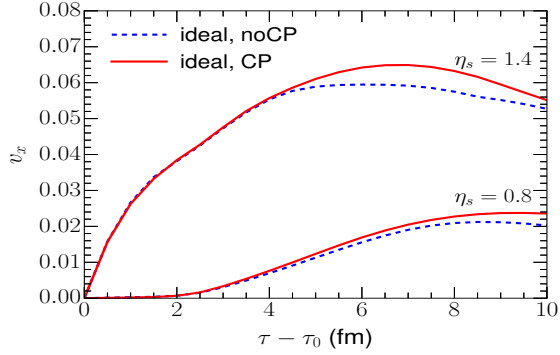} 
\caption{Time evolution for $x$-component of velocity $(v_x)$ of the fluid cell at $x=y=0$ and at different space-time rapidities for Au+Au collision at $\sqrt{s_{NN}}=14.5$ and impact parameter $b=5$ fm.}
\label{fig:vxvst_id}
\end{figure}

\begin{figure*}
\centering
\begin{tabular}{cc}
\includegraphics[height=50mm,width=75mm,angle=0]{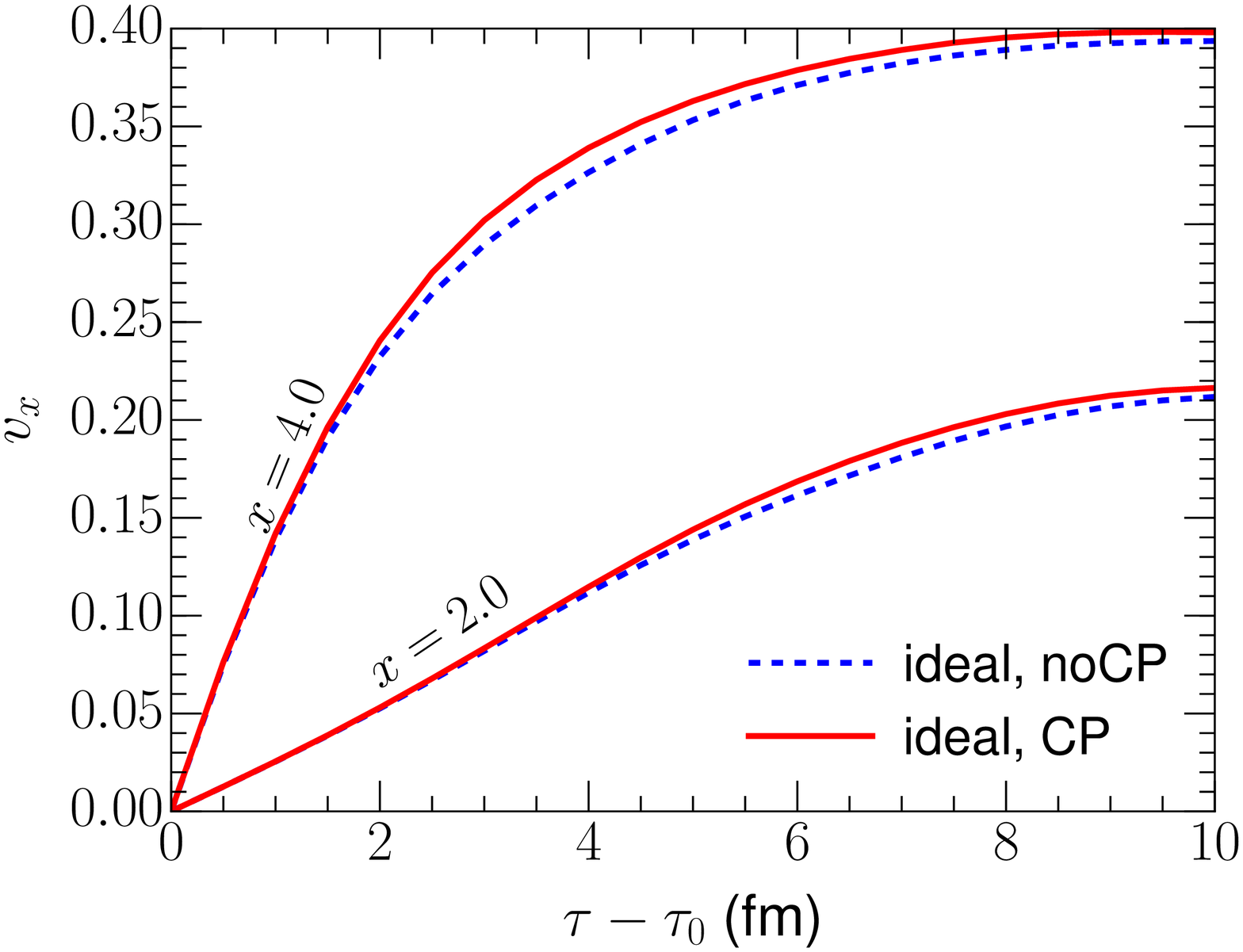} &
\includegraphics[height=50mm,width=75mm,angle=0]{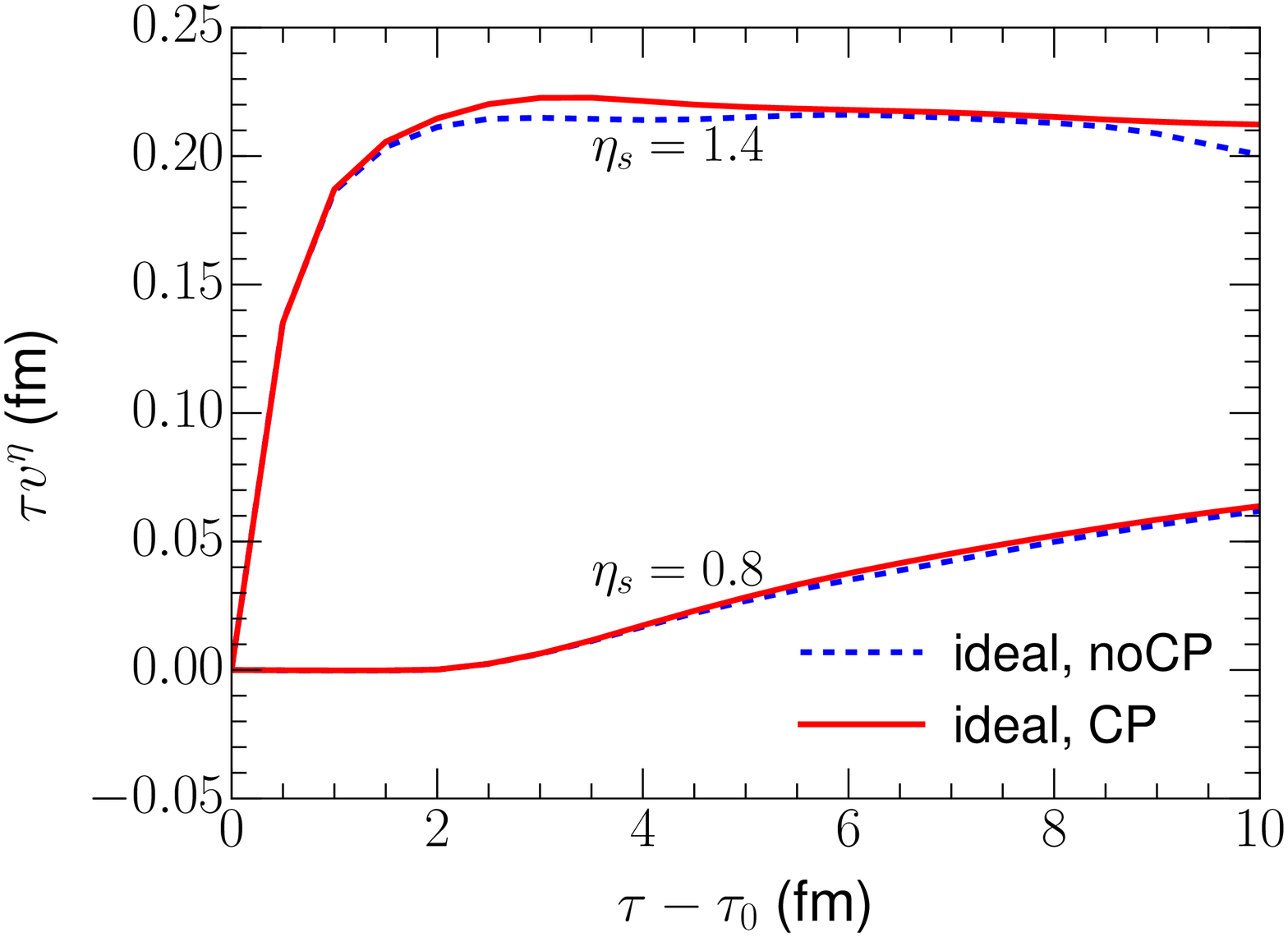}
\end{tabular}
\caption{(left) Time evolution of $v_x$ for the fluid cell at $y=0$, $\eta_s=0$ at different $x$ and (right) time evolution of $v_\eta$ for the fluid cell at $x=y=0$ at different space-time rapidities for Au+Au collision at $\sqrt{s_{NN}}=14.5$ and impact parameter $b=5$ fm.}
\label{fig:vxvzvst_id}
\end{figure*}

In Fig.(\ref{fig:vxvst_id}), we show the time evolution for $v_x$ of a fluid 
cell at different space-time rapidities for Au+Au collision at 
$\sqrt{s_{NN}}=14.5$ and impact parameter $b=5$ fm. The results indicate 
that a gradient of $v_x$ along the $\eta_s$-direction is generated which is due 
to our choice of tilted initial condition.
The presence of CP marginally increases this gradient.
Similarly, the transverse and longitudinal expansions also increases marginally due to the presence of the critical point for ideal hydrodynamics as 
shown in Fig.(\ref{fig:vxvzvst_id}). It should, however, to 
be noted that the profile is mostly monotonic. 

\begin{figure}[H]
\centering
\includegraphics[height=50mm,width=75mm,angle=0]{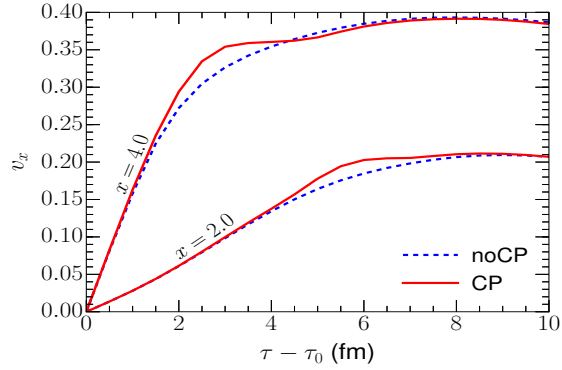} 
\caption{Time evolution of $v_x$ at different $x$ for the fluid cell at $y=$,
$\eta_s=0$ for $\sqrt{s_{NN}}=14.5$ GeV and $b=5$ fm.}
\label{fig:vxvst_diffx_s14p5}
\end{figure}

\begin{figure*}
\centering
\begin{tabular}{cc}
\includegraphics[height=50mm,width=75mm,angle=0]{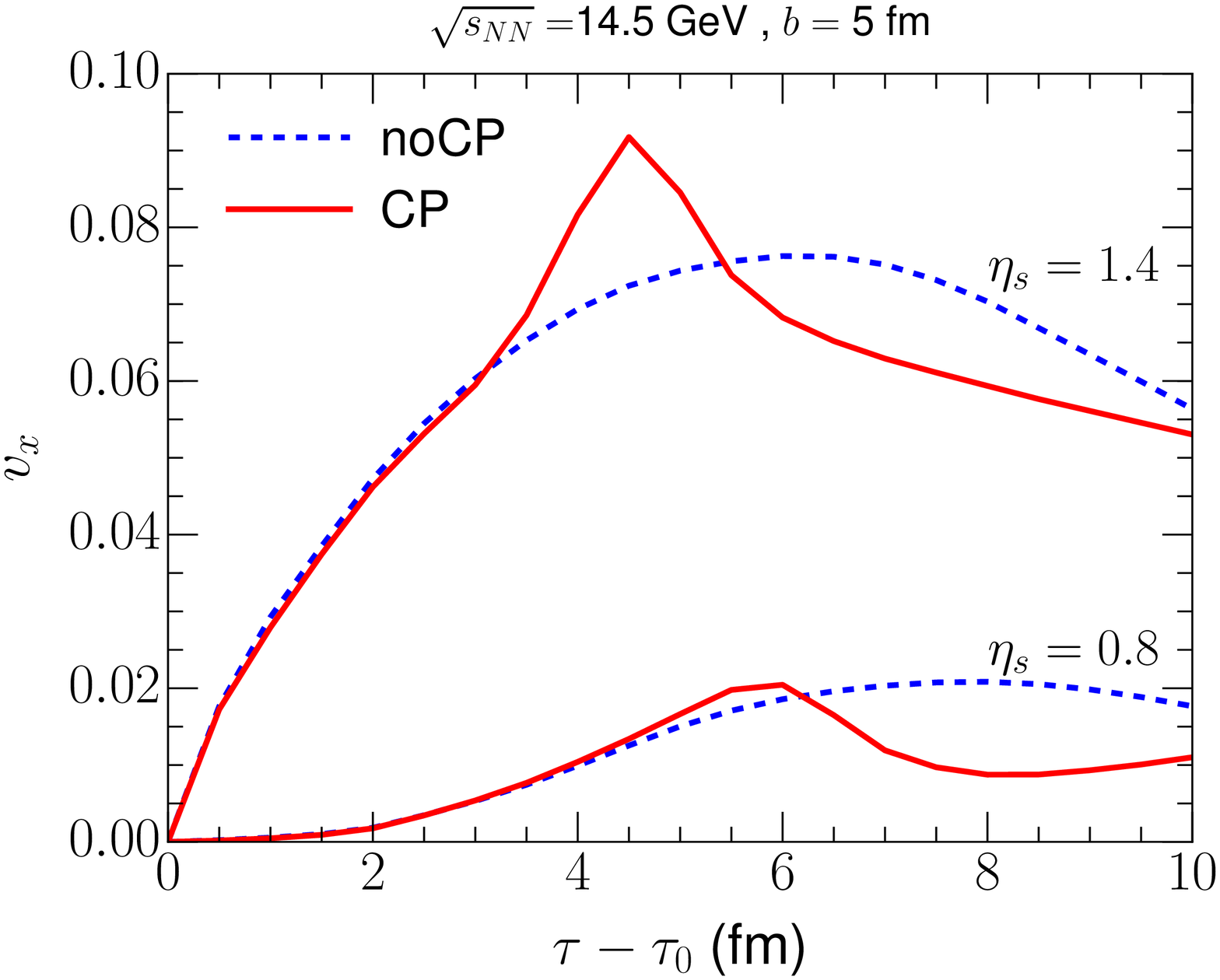} &
\includegraphics[height=50mm,width=75mm,angle=0]{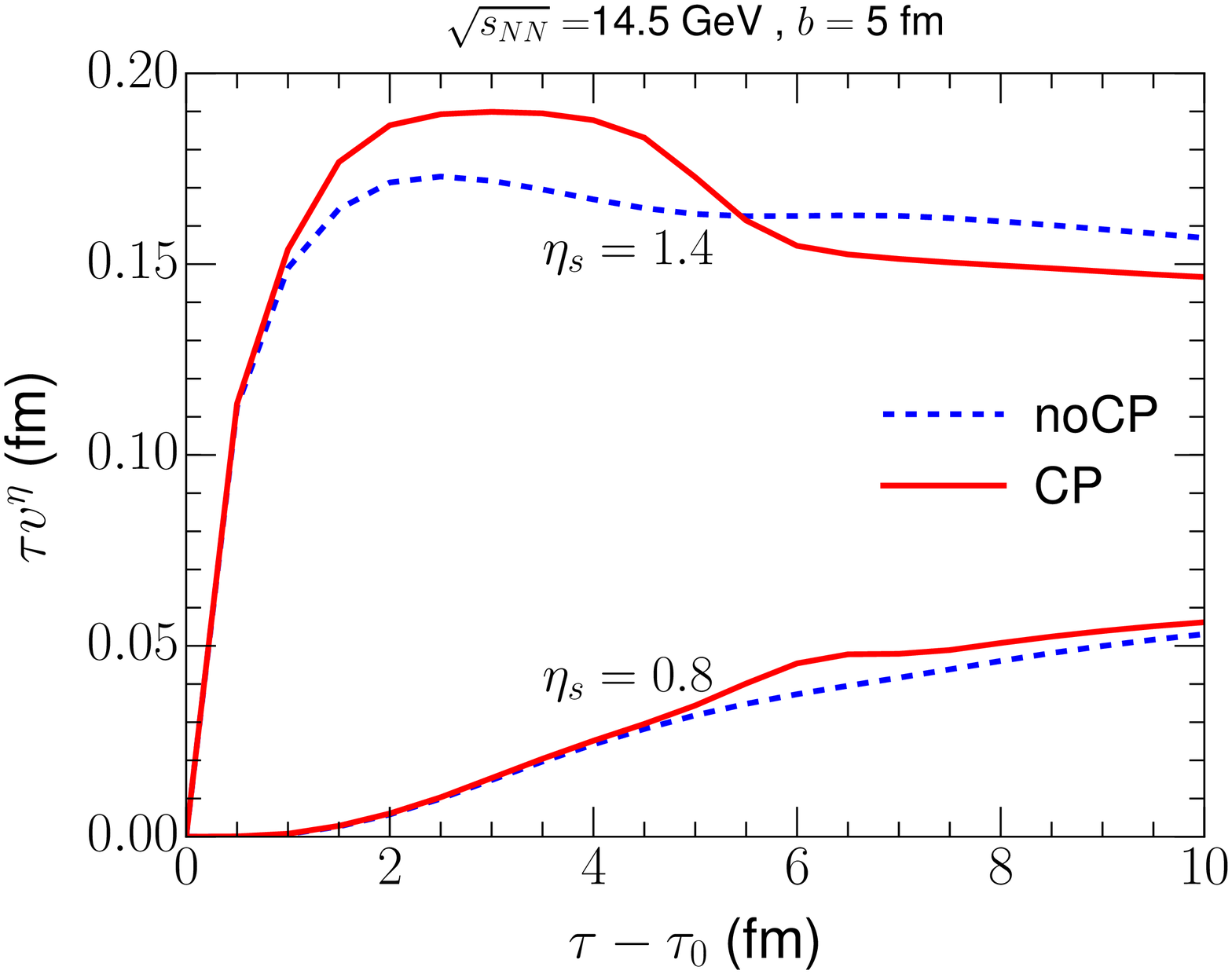} 
\end{tabular}
\caption{Time evolution of $v_x$ at different values of $x$ (left panel) and 
$\tau v_\eta$ at different $\eta_s$ for the fluid cell at $x=y=0$,
for $\sqrt{s_{NN}}=14.5$ GeV and $b=5$ fm.}
\label{fig:vxvzvst_s14p5}
\end{figure*}

\begin{figure*}
\centering
\begin{tabular}{cc}
\includegraphics[height=50mm,width=75mm,angle=0]{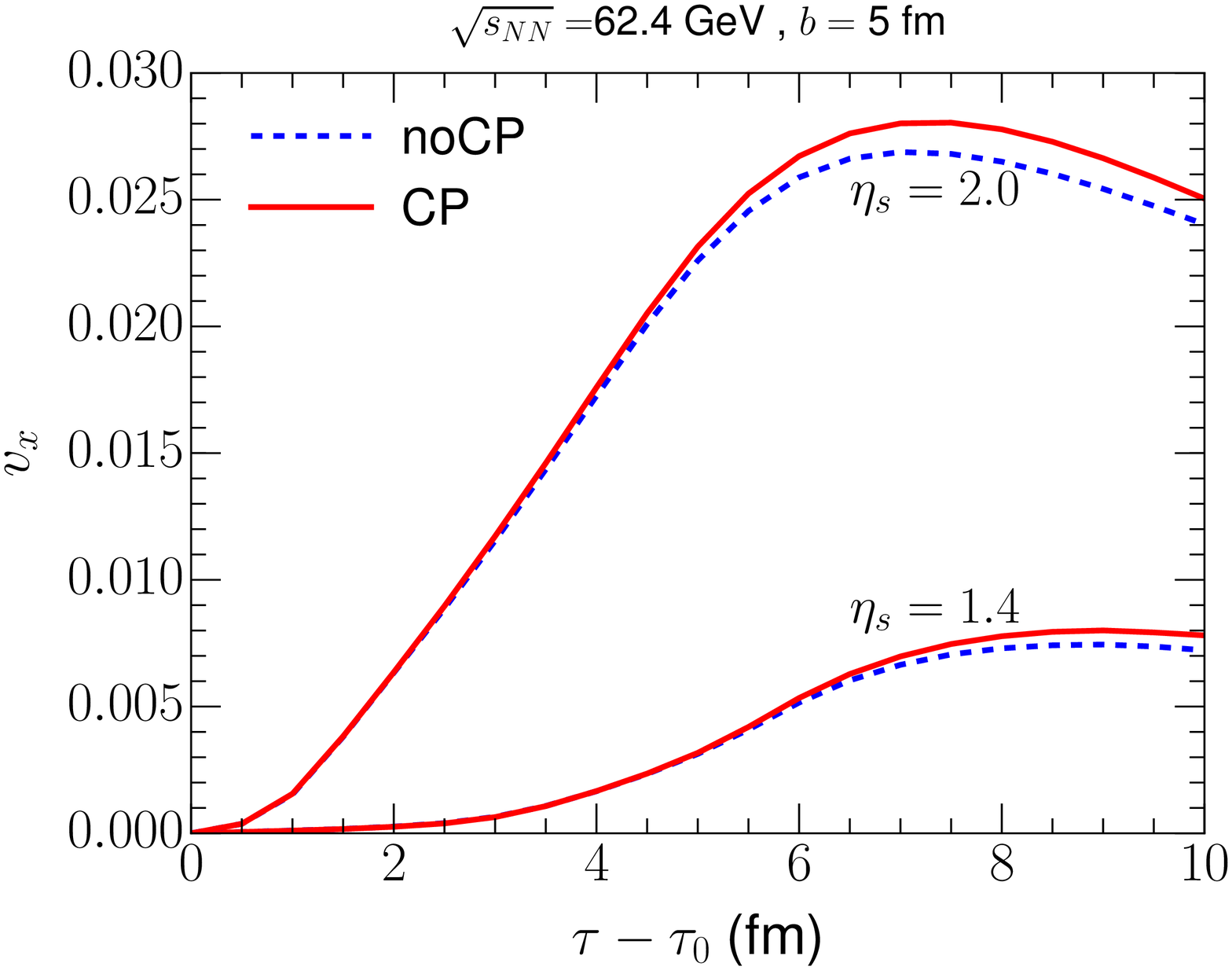} &
\includegraphics[height=50mm,width=75mm,angle=0]{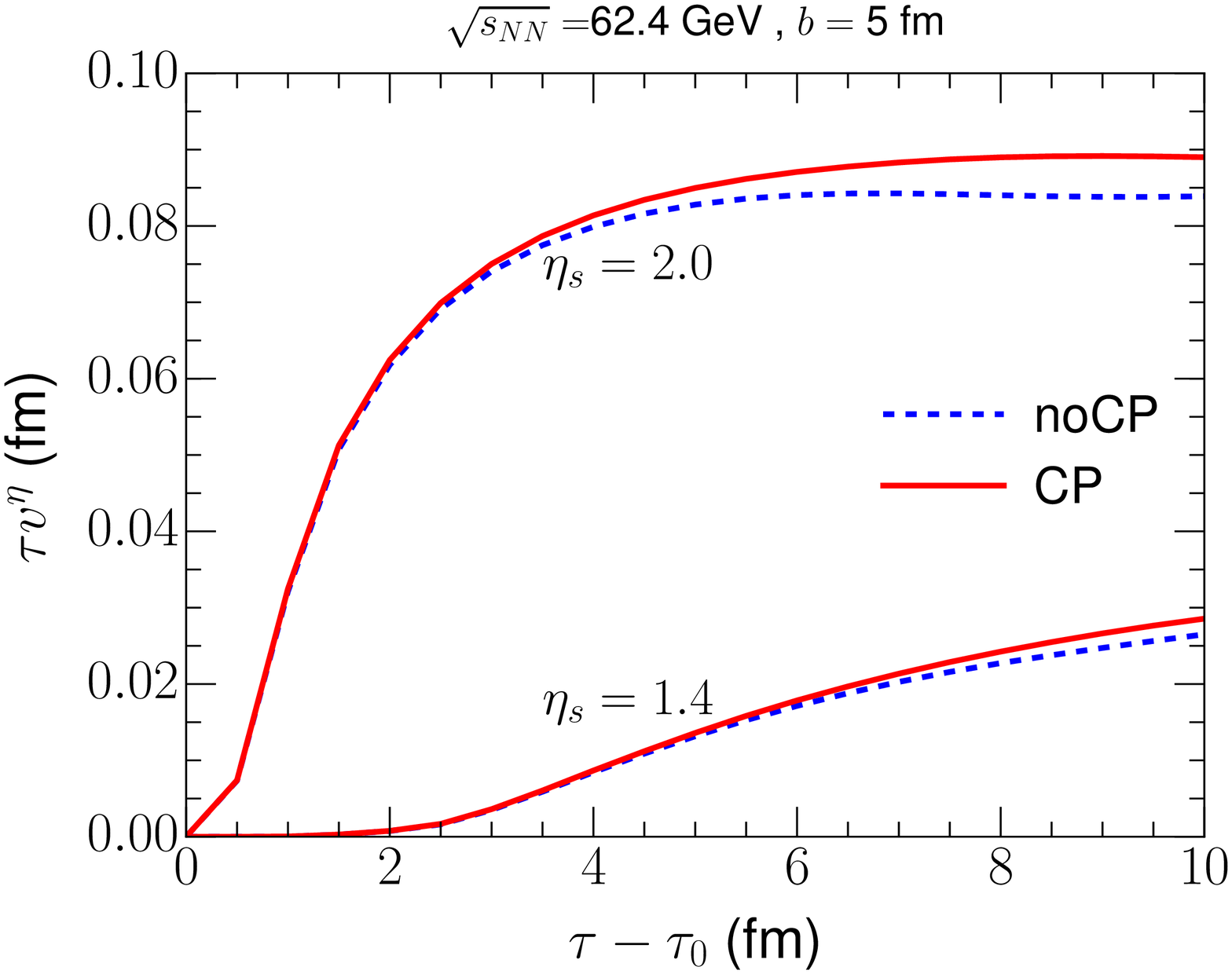} 
\end{tabular}
\caption{Same as Fig.~\ref{fig:vxvzvst_s14p5} for $\sqrt{s_{NN}}=62.5$ GeV.}
\label{fig:vxvzvst_s62p4}
\end{figure*}

Now we discuss results with the inclusion of the viscous effects in the QGP fluid.
As the viscous fireball of QGP expands, the fluid cells towards the boundary, which are closer to the critical region, 
undergo slower expansion due to enhanced viscosity. The expansion of the fluid cells in the bulk  
is not strongly affected by the critical point. This leads to the buildup of matter somewhere in between, 
due to which the expansion results in a non-monotonic profile.
This is reflected in the variation of $v_x$ with $\tau$ 
for different values of $x$ (Figs.\ref{fig:vxvst_diffx_s14p5}) and 
$\eta_s$ (left panel of Fig.~\ref{fig:vxvzvst_s14p5}). 
The variation of $\tau v_\eta$ with $\tau$ for 
different values of $\eta_s$ display similar nature
(right panel of~\ref{fig:vxvzvst_s14p5}). 
The non-monotonicity is prominent for larger values 
of $\eta_s$ which corresponds to the evolution trajectory
closer to the CP (see Fig.~\ref{fig:evln_traj}.  The effect 
of CP on $v_x$ at $\eta_s=1.4$ gives rise to a horn
like structure. We  will check below whether such
structure survives in the space-time integrated 
observables.
It is to be stressed further that as the critical point is approached, these effects get enhanced.
To further confirm this argument, we also 
show the velocity profile for $\sqrt{s_{NN}}=62.4$ GeV in Fig.(\ref{fig:vxvzvst_s62p4}). The trajectories 
in this case are far away from the critical region and thus the effects of EoS and enhanced viscosities are small. 
Therefore, the non-monotonicity observed in the time evolution of $v_x$ and $\tau v_\eta$ can be
attributed to the CP. However, it will be interesting to examine whether such effects of 
CP survives in experimental observables like $p_T$ spectra and various flow coefficients 
which are obtained by integrating over the entire space-time history of the fireball.
This exercise has been carried out below.
 
It is clear that the effects of the CP through the EoS and  
transport coefficients do not affect the velocity much but 
it surely affects the gradients of velocity profile.

\begin{figure*}
\centering
\begin{tabular}{cc}
\includegraphics[height=50mm,width=75mm,angle=0]{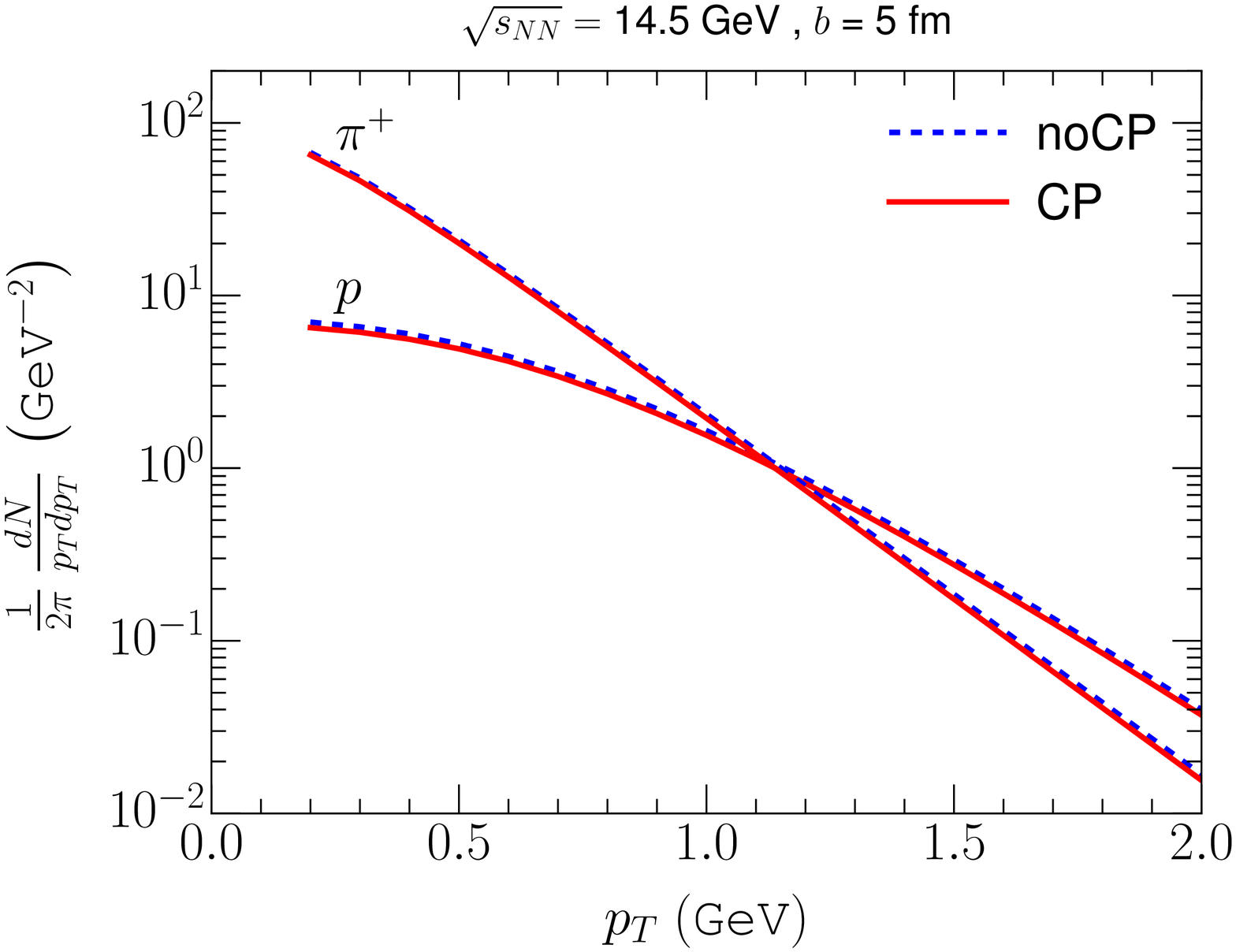} &
\includegraphics[height=50mm,width=75mm,angle=0]{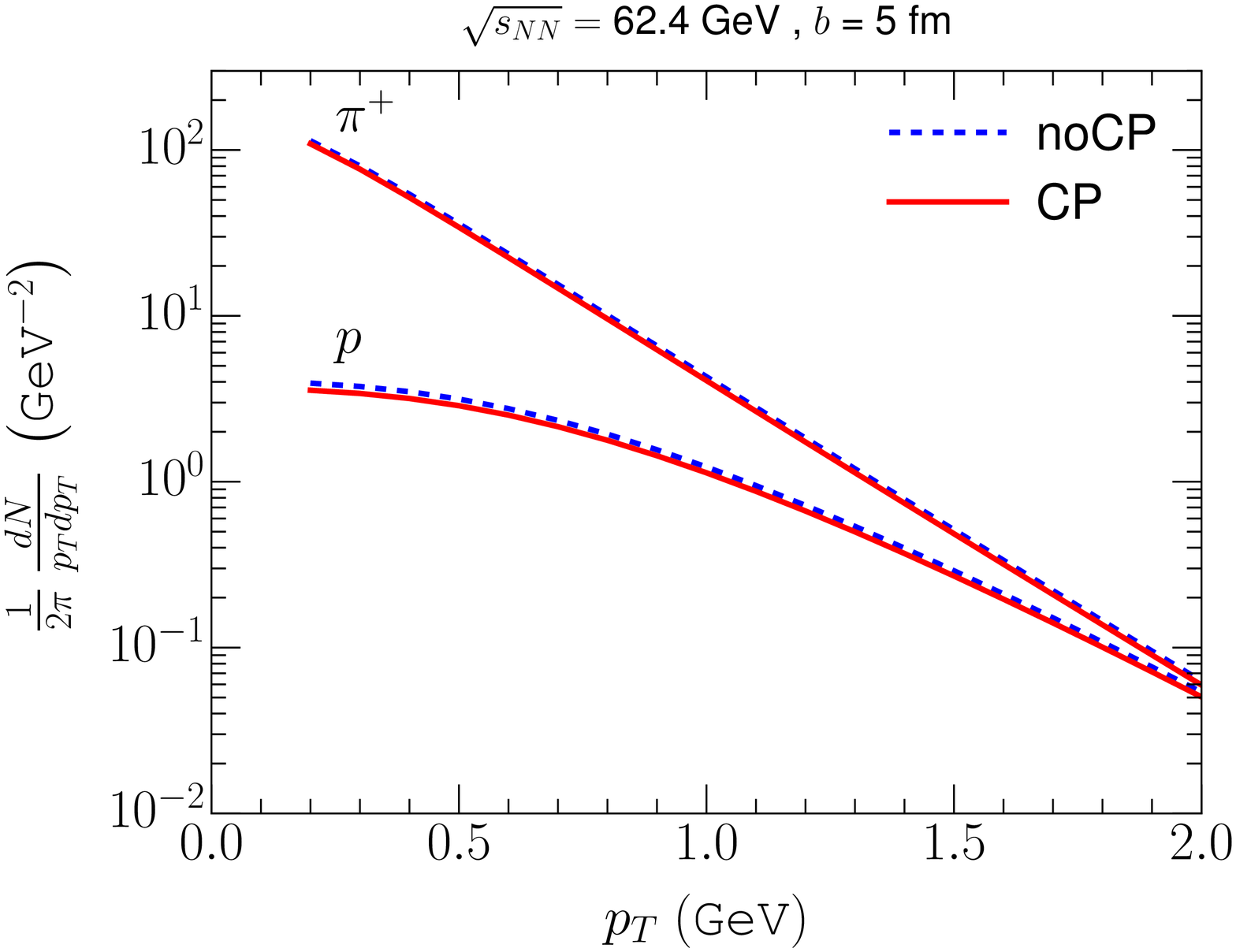} \\
\end{tabular}
\caption{Transverse momentum spectra is obtained by integrating $\phi$ and rapidity $(y)$ between $-1< y < 1$
for $\sqrt{s_{NN}}=14.5$ (left panel) and 62.4 GeV (right panel). }
\label{fig:ptspectra}
\end{figure*}

\begin{figure*}
\centering
\begin{tabular}{cc}
\includegraphics[height=50mm,width=75mm,angle=0]{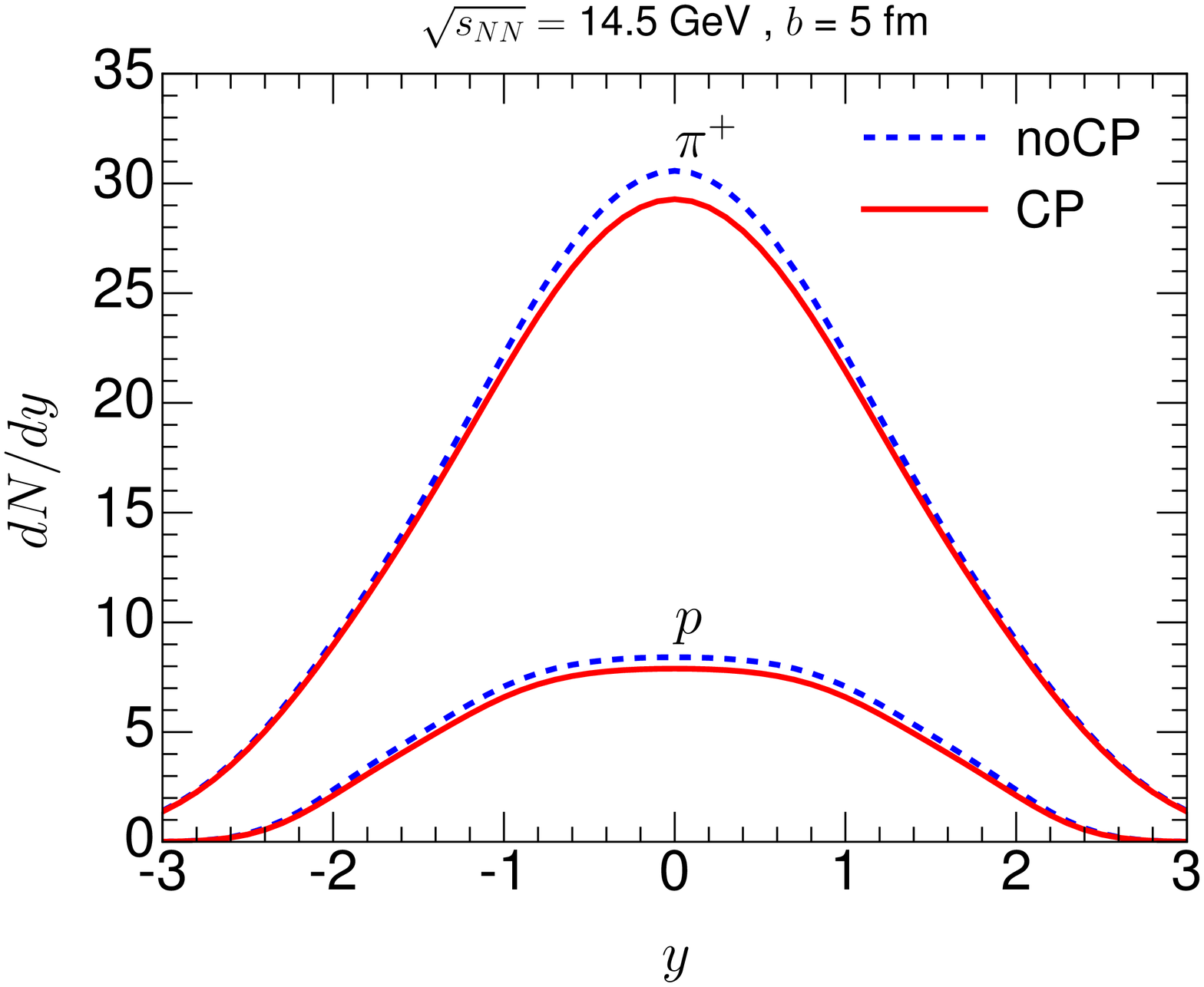} &
\includegraphics[height=50mm,width=75mm,angle=0]{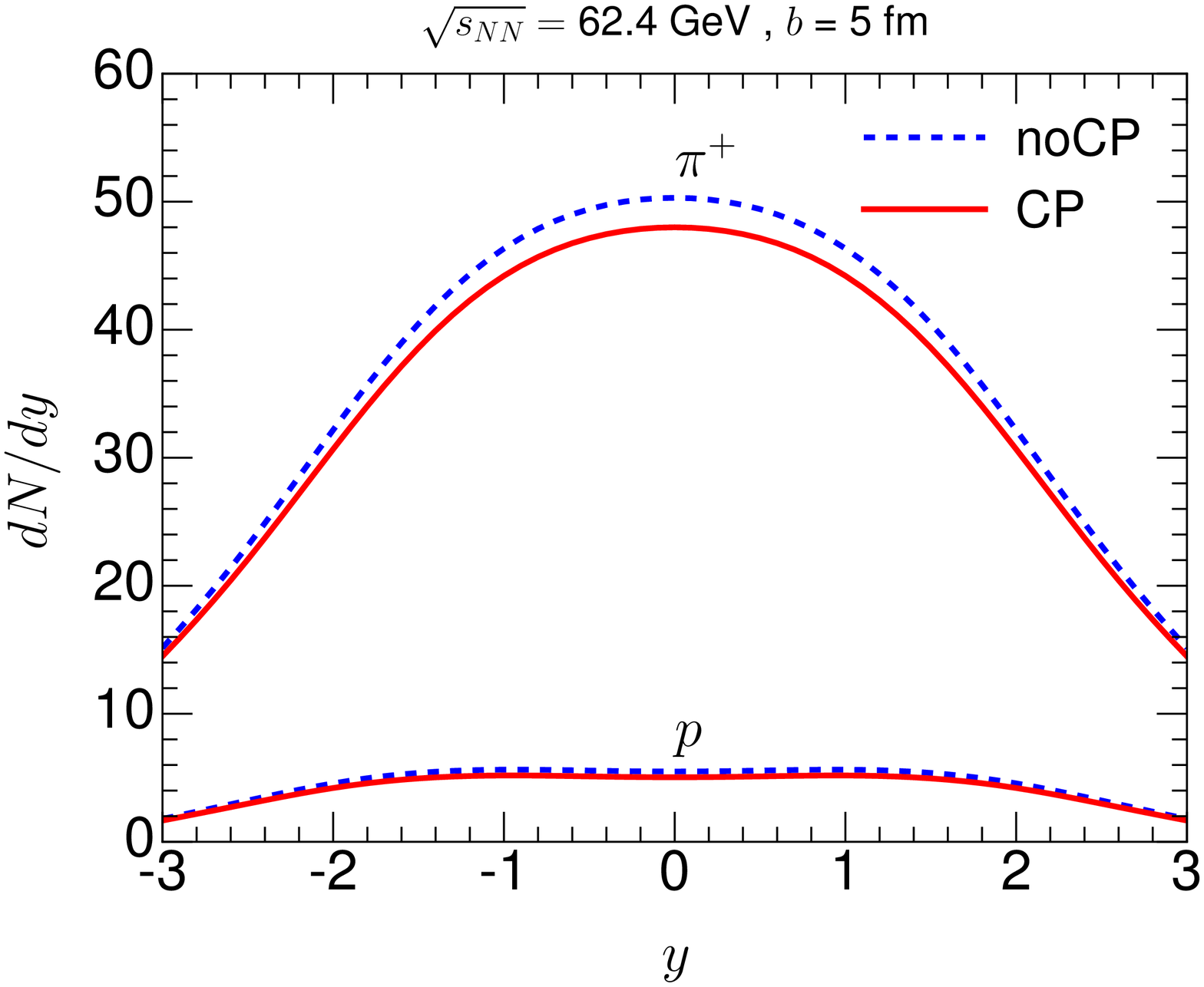} \\
\end{tabular}
\caption{The rapidity distribution of pion and proton  for $\sqrt{s_{NN}}=14.5$ (left panel) and 62.4 GeV (right panel). }
\label{fig:yspectra}
\end{figure*}

We have evaluated  the $p_T$ spectra,  directed  and elliptic flow 
coefficients of pion and proton with the inclusion 
of the effects of CP on the EoS and transport coefficients. 
The $p_T$ spectra of  $\pi^+$ and proton
have been displayed in  Fig.~\ref{fig:ptspectra}) for two 
colliding energies \emph{i.e.} $\sqrt{s_{NN}}=14$ GeV 
(left panel)  62.4 GeV (right panel).  
The rapidity distribution of pion and proton
are shown in Fig.\ref{fig:yspectra} for two colliding energies {\it i.e.} 
$\sqrt{s_{NN}}=14.5$ GeV and
$\sqrt{s_{NN}}=62.4$ GeV, no distinguishable effect of CP is found. 
The effects of CP on the spectra is found to be insignificant
because both the $p_T$ and $y$ distributions  are obtained by integration over
the space-time evolution history of the fireball produced
in these collisions, {\it i.e.} results are superposition of all the temperatures and
densities through which the system  passes, it does not depend  on the
point, ($\mu_c$,\,$T_c$) alone. 

\begin{figure*}
\centering
\begin{tabular}{cc}
\includegraphics[height=50mm,width=75mm,angle=0]{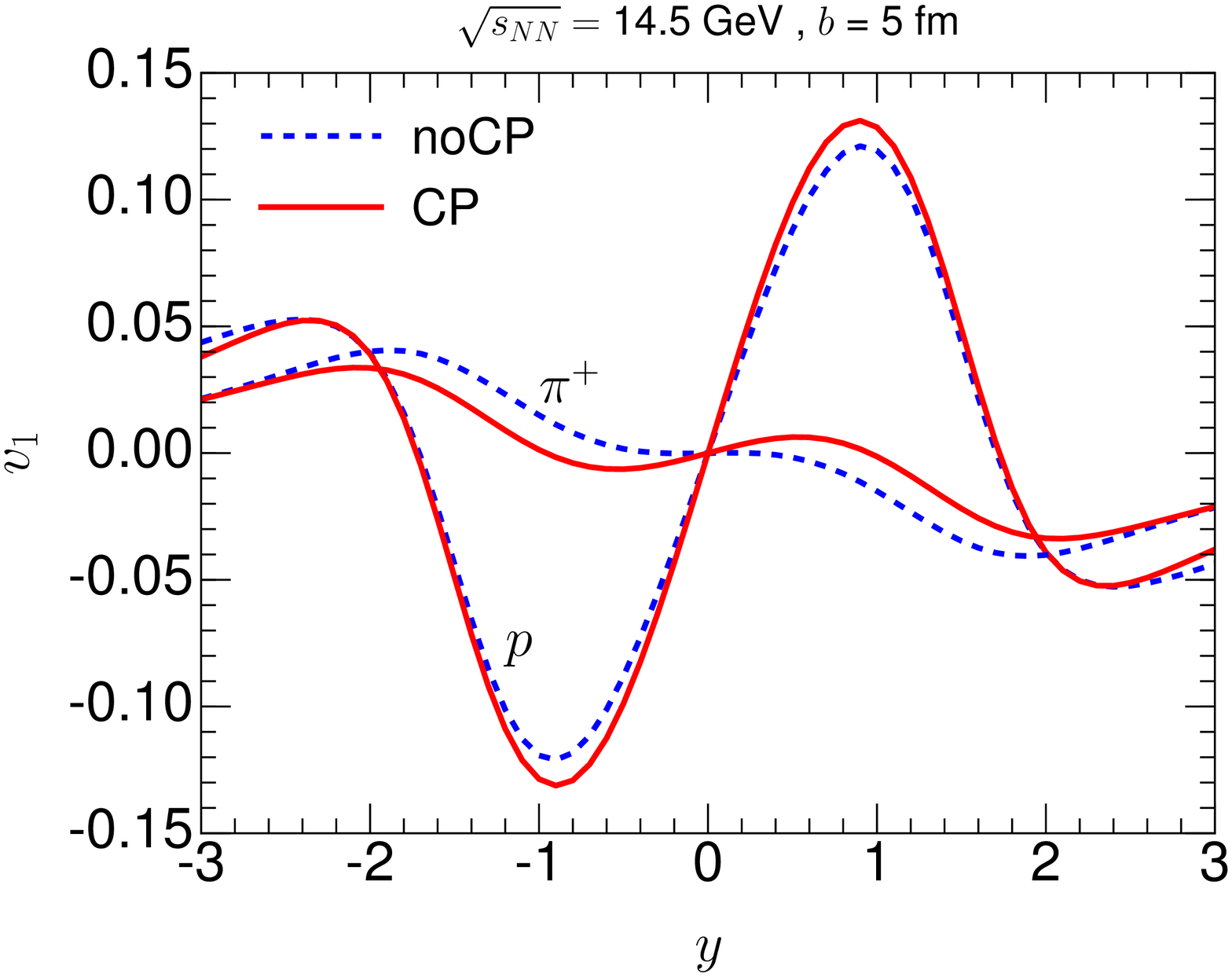} &
\includegraphics[height=50mm,width=75mm,angle=0]{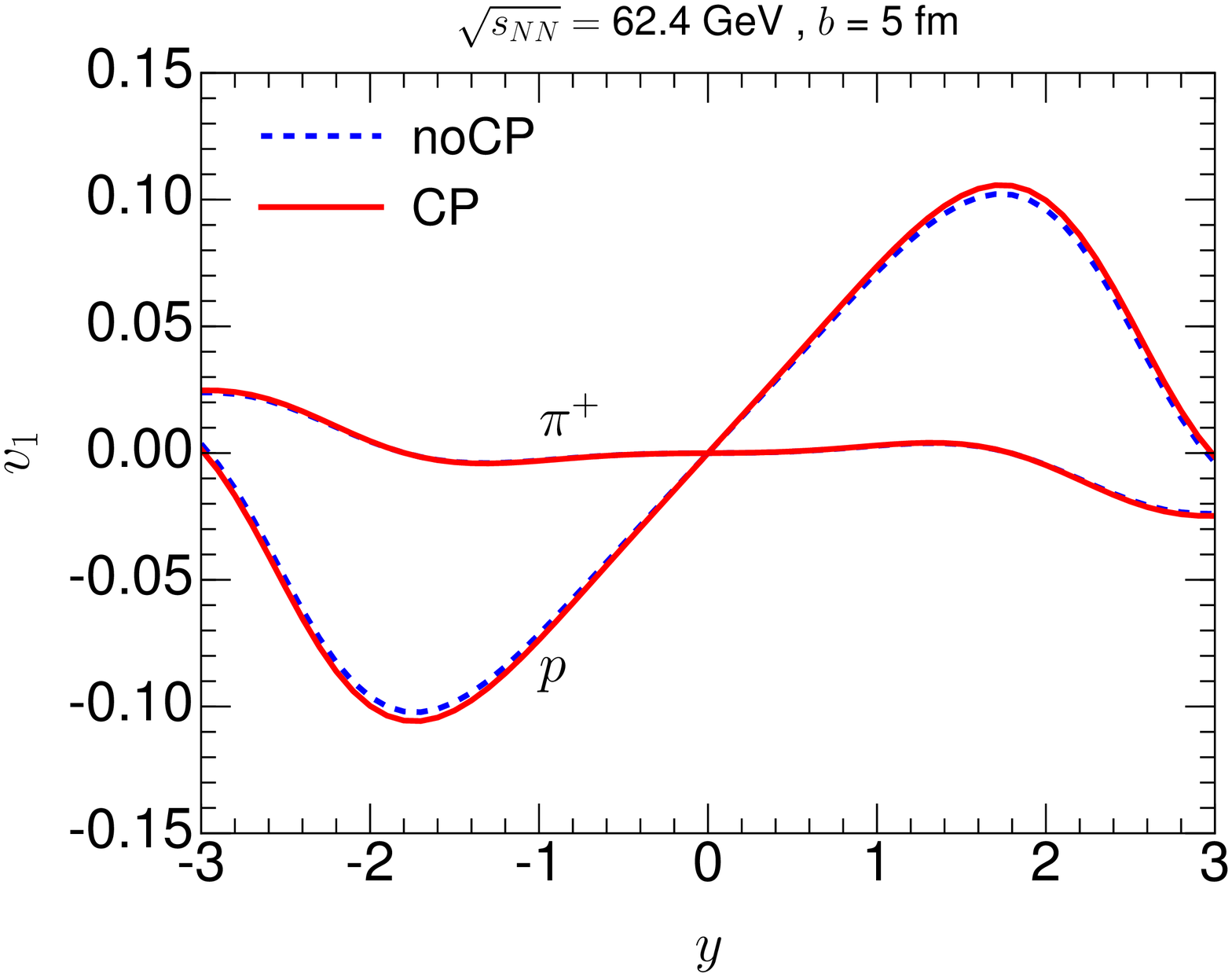} 
\end{tabular}
\caption{Rapidity dependence of directed flow for two colliding energies (14.5 GeV and 62.4 GeV) and for impact parameter $b=5$ fm.}
\label{fig:directedflow}
\end{figure*}

\begin{figure*}
\centering
\begin{tabular}{cc}
\includegraphics[height=50mm,width=75mm,angle=0]{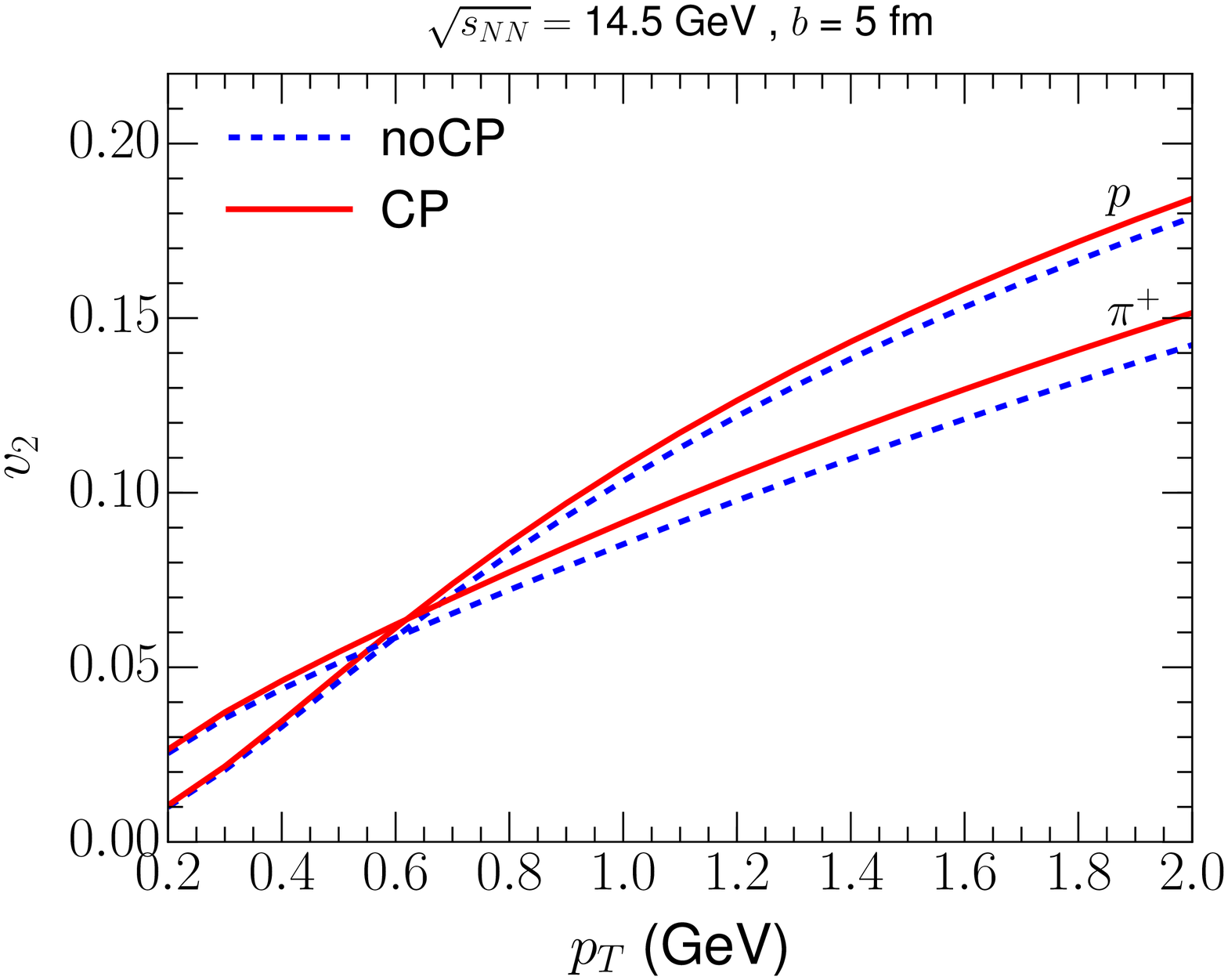} &
\includegraphics[height=50mm,width=75mm,angle=0]{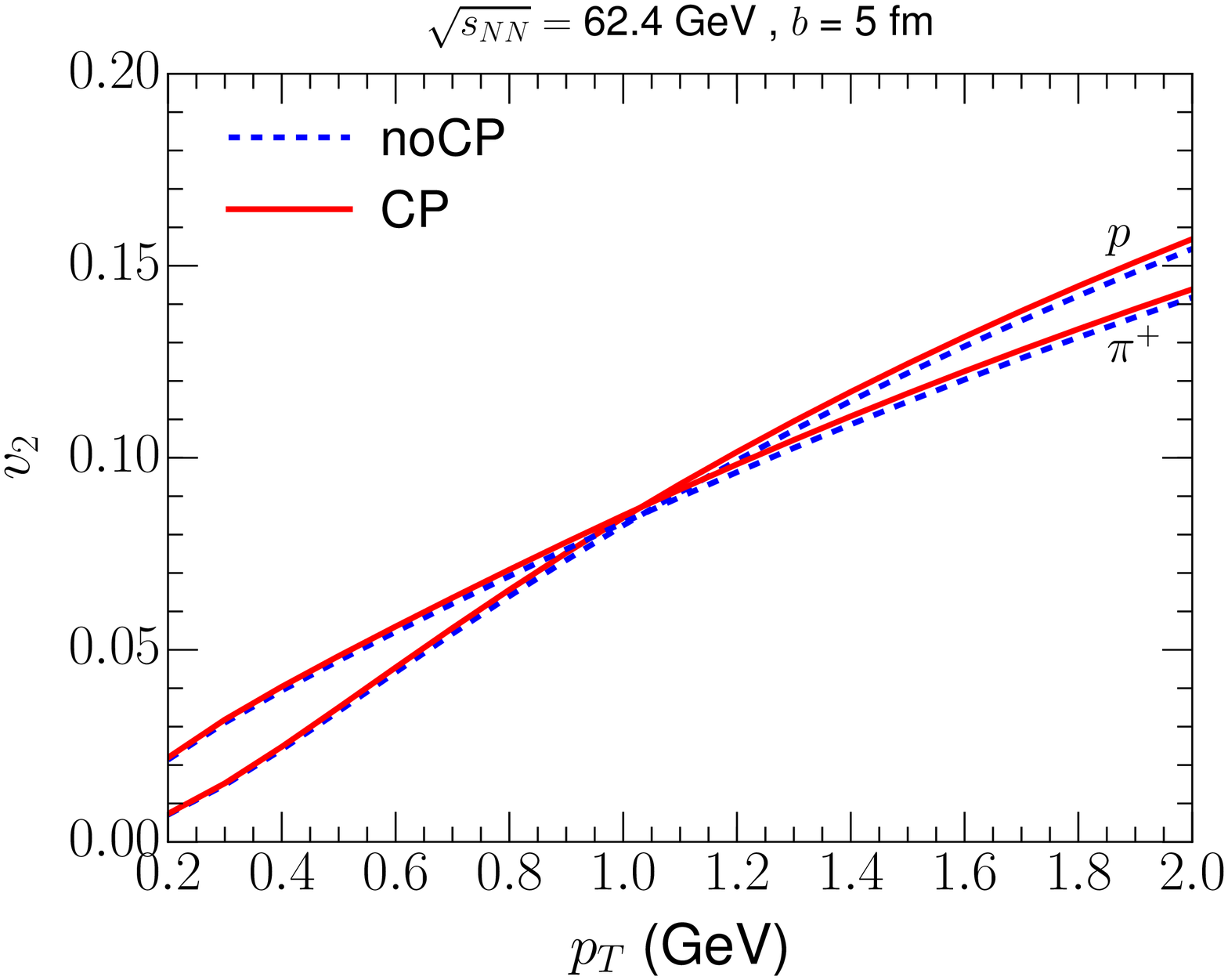} 
\end{tabular}
\caption{$p_T$ dependence of elliptic flow for two colliding energies (14.5 GeV and 62.4 GeV) and for impact parameter $b=5$ fm.}
\label{fig:ellipticflow}
\end{figure*}

The rapidity distribution of $v_1$ of proton and pion are displayed in
Fig.~(\ref{fig:directedflow}) as a function of $y$ for 
$\sqrt{s_{NN}}=14$ GeV  and $62.4$ GeV.  
The effects of CP both on $\pi^+$ and proton are seen
to be insignificant as expected.
The elliptic flow of both protons and $\pi^+$ increases marginally 
around $p_T\sim 2$ GeV compared to the case when there is no CP 
(left panel of Fig.~\ref{fig:ellipticflow}) for $\sqrt{s_{NN}}=14.5$ GeV. 
For $\sqrt{s_{NN}}=62.4$ GeV, there is no shift in $v_2$ due to the
inclusion of CP (right panel of Fig.~\ref{fig:ellipticflow}) which is
expected  because the trajectory in this case remain away from the 
critical domain.

A faster expansion leads to a rapid fall in temperature and chemical potential 
which leads to a slight reduction in the yield of both $\pi^+$ and 
protons as shown in Fig.(\ref{fig:ptspectra}). 
These effects will definitely get enhanced as we approach 
the critical point. 
However, within the current model, the effect is only marginal and might not even be detected experimentally. 
As it has been shown in  Ref.~\cite{Singh:2021yba} that the thermal
vorticity which depends on the velocity gradient of the fluid
is suppressed by the presence of CP. Consequently 
the rapidity dependence of the spin polarization of the $\Lambda$
hyperon was shown to be drastically affected by the CP due 
to the coupling of the spin with the 
thermal vorticity. The CP may be detected by
measuring the rapidity distribution of the spin polarization
by tuning the beam energy. 
Therefore, one expect that those observables which depend on the gradient 
of velocity will be efficient signatures of the CP. 


\section{\label{sec:summary} Summary and Conclusions}
We have developed a computer code to solve the relativistic viscous causal hydrodynamics in (3+1) dimension. 
The effects of the QCD critical point (CP) have been included in the code through the equation of state and scaling 
behaviour of the transport coefficients. We study the evolution of the fireball of quarks and gluons  formed
at different values of $\mu_B$ and $T$ corresponding to  various values of $\sqrt{s_{NN}}$.
The $p_T$ spectra, directed and elliptic flow coefficients of pions and protons  have
been evaluated to understand the effects of CP on these quantities. We  find that the integration over
the entire space-time history of the fireball mostly wipe out the effects of CP on the spectra and flow coefficients
which indicates that the detection of CP by using the hadronic spectra may not be useful.
As the CP has the potential to substantially alter the rapidity distribution of spin polarization of hadrons,
the measurement of spin polarization as a function of rapidity can be considered as an efficient tool to detect the 
CP ~\cite{Singh:2021yba}. 

A few comments on the validity of hydrodynamics near the CP are
in order at this point.
It is well-known that the fluid dynamics becomes invalid,
because the fluctuating modes near the CP do not relax faster than
the time scale of changes in slow/conserved variables due to which 
the local thermal equilibrium can not be maintained.
However, the results obtained in this work may be considered as
reasonable  because  of the following reasons.
Firstly, the  validity of the fluid dynamics can
be extended by adding a scalar field representing the slow non-hydrodynamic modes
connected to  the relaxation  rate of the critical fluctuation
(see ~\cite{Stephanov:2017wlw} and  \cite{stephanov2018} for details).
In a simple picture describing the evolution of a system near the CP,
it has been explicitly shown that the modes associated with the scalar fields
lags behind the hydrodynamic modes resulting in back reactions on the
hydrodynamic variables, however, it was found  
that the  back reaction has negligible effects on the 
hydrodynamic variables~\cite{rajagopal}.
Secondly, one may recall that if a system
is not too close to CP  then hydrodynamics can
still be applied in a domain around the CP~\cite{Stanley}.

\bibliographystyle{apsrev4-2}
\bibliography{refs}
\end{document}